\documentstyle[12pt]{article}
\renewcommand{\theequation}{\Roman{section}.\arabic{equation}\alph{subeq}}
\renewcommand{\thefigure}{\Roman{section}.\arabic{figure}}
\newcounter{subeq}
\newcommand{\eqnreset}{\setcounter{subeq}{0}}
\newcommand{\eqna}{\setcounter{subeq}{1}}
\newcommand{\eqnb}{\addtocounter{equation}{-1}{\setcounter{subeq}{2}}}
\newcommand{\eqnc}{\addtocounter{equation}{-1}{\setcounter{subeq}{3}}}

\newcommand{\bgfrac}[2]{\mbox{$\displaystyle{\frac{#1}{#2}}$}}

\newcommand{\Z}{\mbox{$Z^0$}}

\newcommand{\be}{\begin{equation}}
\newcommand{\ee}{\end{equation}}
\newcommand{\bea}{\begin{eqnarray}}
\newcommand{\eea}{\end{eqnarray}\eqnreset}
\begin{document}

%
%
\renewcommand{\theequation}{\arabic{equation}\alph{subeq}}
\renewcommand{\thefigure}{\arabic{figure}}
\renewcommand{\thetable}{\arabic{table}}
\renewcommand{\theenumi}{\roman{enumi}.)}
\renewcommand{\labelenumi}{\roman{enumi}.)}

\def\be{\begin{equation}}
\def\ee{\end{equation}}
\def\Z{$Z^{\circ}$ }
\newcommand{\si}{$\sin^2 \Theta_W$}
\def \e{$|K_{L, S} (0) > $}
\def \h{$ H_{\rm eff}$}

\title{Standard Model Expectations for CP Violation\thanks{Lectures presented
at the School in Particle Physics and Cosmology, Puri (Orissa), India, January
1993.  To appear in the Proceedings of the School.}}

\author{R. D. Peccei \\ Department of Physics \\ University of
California, Los Angeles, California 90024-1547}

\maketitle

\begin{abstract}
I review the predictions and expectations of the CKM model for CP violation in
both the $K^0-\bar K^0$ and $B^0-\bar B^0$ systems.  A brief discussion of CP
violation in charged $K$- and $B$-decays is also included, as well as some
remarks on the electric dipole moments of the neutron and the electron.
\end{abstract}

\pagebreak

\section{Prologue}

These notes contain a summary of the lectures I delivered at the School on
Particle Physics and Cosmology held in Puri (Orissa), India, in January 1993.
To keep the length of this manuscript manageable, I have not included here two
topics which I discussed in Puri: the strong CP problem and invisible axions
and CP violations and baryogenesis.  The first topic is reviewed by me already
rather comprehensively in Cecilia Jarlskog's monograph on CP violation
\cite{Cecilia} and it did not seem reasonable to repeat much of this discussion
here again.  The second topic, by itself, seemed somewhat disconnected from the
rest of the other material and, regretfully, I decided to leave it out.  For a
discussion of some general issues involved in CP violation and baryogenesis,
the interested reader is referred to my contribution to {\bf 25 Years of CP
Violation} \cite{RP}.  For a more thorough treatment of baryogenesis at the
electroweak scale, the recent lectures of Shaposhnikov \cite{Misha}
are highly recommended, as is the rapporteur talk of A. Cohen at the
PASCOS/Texas '92 Conference \cite{AC}.

\section{Summary of Experimental Information on CP Violation}

I begin these lectures by reviewing what we know experimentally about CP
violation.  At present, we have only observed CP violation in the $K^0 -
{\bar{K}}^0$ complex.  However, important information on CP violation
can also be deduced from the existing bounds on the electric dipole
moment of the neutron and that of the electron, as well as from the
ratio of baryons to photons in the universe.

\subsection{Measurement in the Neutral Kaon System}

Christensen, Cronin, Fitch and Turlay's \cite{CCFT} observation of the
decay $K_L \to \pi^+ \pi^-$ occurred nearly 30 years ago. Since that
time many sophisticated experiments have continued to probe for CP
violation in the $K^0 - {\bar{K}}^0$ complex.  At present, all positive
signals of CP violation can be summarized in terms of 5 measured
parameters:  two complex amplitude ratios $(\eta_{+ -}~{\rm and}~
\eta_{00})$ and the semileptonic rate differences $(A_{K_{L}})$.  More
precisely, the quantities measured are:
\[
\eta_{+ -} = \frac{A (K_L \to \pi^+ \pi^-)}{A(K_S \to \pi^+ \pi^-)}
\equiv ~ \vert \eta_{+-} \vert~ e^{i \phi_{+ -}} \equiv \epsilon +
\epsilon^{\prime} \]
\[\eta_{00} = \frac{A(K_L \to \pi^0 \pi^0)}{A(K_S \to \pi^0 \pi^0)}
\equiv \vert \eta_{00} \vert~e^{i \phi_{00}} \equiv \epsilon - 2
\epsilon^{\prime}\]
\be A_{K_{L}} = \frac{\Gamma(K_L \to \pi^- \ell^+ \nu_{\ell})~ -~
\Gamma (K_L \to
\pi^+ \ell^- {\bar{\nu}}_\ell )}{\Gamma (K_L \to \pi^- \ell^+ \nu_{\ell})~ +~
\Gamma (K_L \to \pi^+ \ell^- {\bar{\nu}}_{\ell})}
\ee
To a very good approximation, one finds that
\be
\eta_{+-} ~\simeq~ \eta_{00}~~.
\ee
(and therefore $\epsilon^{\prime} \ll \epsilon$) and that
\be
A_{K_{L}} ~ \simeq~ 2~ Re \eta_{+-}~~;~~ \phi_{+-}~\simeq~ \phi_{00}
\simeq \pi/4~~.
\ee
The first result above, as we shall see, tells one that CP violation in
the neutral Kaon system is mostly due to mixing.  In view of the fact
that $\epsilon^{\prime}$ is much less than $\epsilon$, the latter two
results provide tests of CPT conservation in neutral Kaon decays.
Again, this will be elaborated upon below.

In more detail, there is at
the moment conflicting evidence regarding $\epsilon^{\prime}/\epsilon$,
with the NA31 experiment at CERN reporting a $3 \sigma$ positive signal
for $Re~ \epsilon^{\prime}/\epsilon$, but the E731 Fermilab experiment
still finding a signal consistent with zero:

\bea
Re \frac{\epsilon^{\prime}}{\epsilon} = \left\{ \begin{array}{cc}
(23 \pm 7) \times 10^{-4} & \mbox{\protect \cite{NA31}} \\
(7.4 \pm 5.9) \times 10^{-4} & \mbox{\protect \cite{EZ31}}
\end{array}\right.
\eea
The imaginary part of this ratio, measured through the phase difference
between $\phi_{+-}$ and $\phi_{00}$, to the accuracy with which it is
measured at present, is also consistent with zero.  One finds
\bea
3~ Im~ \epsilon^{\prime}/\epsilon ~ = \phi_{+-} - \phi_{00} = \left\{
\begin{array}{cc} ( - 0.2 \pm 2.6 \pm 1.2)^0 & \mbox {\protect
\cite{NA32}} \\
(1.6 \pm 1.0 \pm 0.7)^0 & \mbox{\protect \cite{EZ31}} \end{array}\right.
\eea

In addition to the above, the compilation of the Particle
Data Group \cite{PDG}, gives
\bea \eqna
\vert \eta_{+-} \vert & = (2.268 \pm 0.023) \times 10^{-3} \\ \eqnb
A_{K_{L}} & = (3.27 \pm 0.12) \times 10^{-3}
\eea
and
\bea \eqnc
\phi_{+-} & = (46 \pm 1.2)^0~~.
\eea
However, this last number has been brought into question by the recent
reanalysis performed by the E731 collaboration.  Using a somewhat
smaller value of the $K_L - K_S$ mass difference $\Delta m$, the E731
reanalysis
yields,
instead of the PDG value above \cite{PDG}, the average value \cite{EZ31}
\be
\phi_{+-} = (42.8 \pm 1.1)^0
\ee
This value is in much better accord with what one predicts from CPT
conservation \cite{RDP} where one expects\footnote{Here \protect
$\Gamma_S$ and $\Gamma_L$ are the widths of
$K_S$ and $K_L$, respectively.}
\be
\phi_{+-} \simeq \phi_{sw} = \tan^{-1}
\frac{ 2 \Delta m}{\Gamma_S - \Gamma_L} = (43.4 \pm 0.2)^0
\ee
\subsection{Bounds on Electric Dipole Moments}

Landau \cite{Landau} was the first to point out that, for an elementary
particle, having an electric dipole moment violates both $P$ and $T$.  If
one assumes that CPT is conserved, as is expected from the CPT theorem
\cite{CPT}, then the presence of an electric dipole moment would also
signal CP violation.

A simple argument to see why an electric dipole moment $\vec{d}$
violates both $P$ and $T$ is as follows \cite{Nir}.  Since $\vec{d}$ is
a 3-vector, and measures a static property of an elementary particle,
it must be proportional to the only other 3-vector in the problem - the
angular momentum $\vec{J}$.  Thus
\be
\vec{d} = d \vec{J}~~.
\ee
However, $\vec{d}$ is odd under $P$, while $\vec{J} \to \vec{J}$ under
parity.  Hence, if $P$ is conserved, the constant $d$ must vanish.
Similarly, $\vec{d}$ is even under $T$, but $\vec{J} \to - \vec{J}$ under
time reversal transformations.  Hence again, unless $T$ is violated, $d$
must vanish.

Experimentally, there are strong limits on the electric dipole moments
of both the neutron and the electron.  From the Particle Data Group
\cite{PDG} one has
\bea
d_n &<&   1.2 \times 10^{-25}{\rm ecm}~~~~  (95\%~ C.L.) \nonumber \\
d_e  &=&  (-0.3 \pm 0.8) \times 10^{-26}{\rm ecm}
\eea
\subsection{Astrophysical and Cosmological Information on CP Violation}

There is a nice cosmological argument, due to Zeldovich, Kobzarev and
Okun \cite{ZKO} which strongly suggests that the violation of CP seen
experimentally in the neutral Kaon system must signal an {\bf explicit}
violation of this symmetry in the Lagrangian of the theory, rather than a
spontaneous breaking of CP by the vacuum.  If CP were to be a
spontaneously broken symmetry, with a breaking scale $v_{\not cp}$, one
would expect that at temperatures below $T^{\ast} \sim v_{\not cp}$ different
CP domains would form in the universe.  These domains would be separated
from each other by domain walls of typical surface energy density
\be
\sigma \sim T^{\ast 3} \sim v^3_{\not cp}~~.
\ee
However, unless $v_{\not cp}$ is extremely small, which is not sensible
since one expects that $v_{\not cp}$ be at least as big as the scale of
electroweak breaking $v \sim 250~GeV$, the energy in these domain
walls today would far exceed the closure density of the universe.
\footnote{This problem can be avoided in inflationary universe scenarios
if $v_{\not cp}$ is greater than the scale when inflation takes place.
However, then one has to worry about transmitting spontaneous CP
violation occuring at these very large scales to the neutral Kaon sector
\protect\cite{RP}.}  One has
\be
\rho_{\rm wall} \sim \sigma T \sim v^3_{\not cp} T \sim 10^{-7} \left(
\frac{v_{\not cp}}{TeV} \right)^3 ~GeV^{-4},
\ee
to be compared to the closure density of the universe today
\be
\rho_{\rm closure} \sim 10^{-46}~GeV^{-4}~~.
\ee

There is a second place where cosmology and astrophysics have a bearing
on the issue of CP violation, related to the ratio $\eta$ of
baryons to photons in the universe today.  This ratio is rather well
determined from the study of the primordial abundances of the light
elements produced in nucleosynthesis and one finds\cite{G. Steigman}
\be
3.7 \times 10^{-10} < \eta < 4.0 \times 10^{-10}~~.
\ee
If the universe was symmetric in the number of baryons and antibaryons
at temperatures above a few GeV, then from subsequent annihilations $(p +
\bar{p} \to 2 \gamma )$ one would expect a ratio $\eta$ only of order
$\eta \sim 10^{-18} \cite{KT}$.  Therefore, this ratio must reflect a
primordial baryon-antibaryon asymmetry.  That is,
\be
\eta = \frac{n_B - n_{\bar{B}}}{n_{\gamma}}~~.
\ee

It is possible - but very difficult to conceive physically - that $\eta$
is an initial condition for our universe.  If so, one learns nothing
from this number.  However, it is much more reasonable that $\eta$ be
produced dynamically in the course of the evolution of the universe.
In this case, as Sakharov \cite{Sakharov} was the first to point out,
the dynamics necessarily must involve CP violating
phenomena.  Thus the ratio $\eta$ itself is also a measure of CP
violation.  The relation of $\eta$ with the CP violating phenomena seen in the
kaon system is, however, far from direct \cite{RP}, even for baryogenesis
produced at the electroweak scale. \cite{Misha}

\section{CKM Paradigm}

There is an important consequence that follows from assuming that the
observed CP violation is due to explicit CP breaking in the underlying
Lagrangian.  Namely, if we want this Lagrangian to be renormalizable,
then once CP is no longer a symmetry it follows that {\bf all} parameters
of this Lagrangian that can be complex must be so.  Otherwise, one could
not absorb potential infinities into appropriately complex counter
terms.  In the standard model of the strong and electroweak
interactions, the gauge sector is necessarily real so no CP phases can
enter through the gauge coupling constants.\footnote{There can be CP
violation associated with the gauge interactions as a result of the
presence of non-trivial vacuum angles $\theta$. This matter is not
very germane to the present discussion, and will not be examined here further.
For a discussion, see, for example\cite{Cecilia}.}.  It
follows, therefore, that any CP violation in this model must arise as a
result of interactions in the Higgs sector.

If one has only one complex doublet of Higgs fields $\Phi$, as is
generally assumed in the simplest version of the standard electroweak
theory, then any CP violating phases can only appear in the Yukawa
interactions because, by Hermiticity, the Higgs potential has only real
parameters:
\be
V = \lambda (\Phi^\dagger \Phi - \frac{v^2}{2} )^2~~,
\ee
with $\lambda$ and $v^2$ real.  However, when one has more than one
Higgs doublet, CP violating phases can also enter in the pure Higgs
sector.  For instance, with two Higgs doublets, $\Phi_1$, and $\Phi_2$,
one can have a complex mass term
\be
{\cal{L}} = - \mu^2 \Phi_1 \Phi_2 - (\mu^2)^{\ast} \Phi^{\dagger}_1
\Phi^{\dagger}_2~.
\ee

It is often useful in confronting novel phenomena to describe them in a
model with the minimum number of free parameters.  This is precisely
what occurs with CP violation in the standard electroweak model with
only one Higgs doublet, in the case in which there are three generations
of quarks (and leptons).  In this physically relevant circumstance, all
CP violating phenomena are traceable to a single phase arising from the
Yukawa couplings of quarks to the doublet Higgs boson.  This is the well
known Cabibbo Kobayashi Maskawa (CKM) paradigm \cite{CKM}.  I shall, in
what follows,
on
the whole, concentrate on the prediction of this model.
There may well be further CP violating phases in nature
besides the CKM phase.  However, since we know that the CKM phase must exist
by the renormalizability of the standard model, it seems reasonable to
see first if this phase indeed can explain all the observed CP violating
phenomena in the Kaon system.

\subsection{Counting CP phases in the Standard Model}

If CP is not a good symmetry, the Yukawa couplings of quarks to the
Higgs doublet $\Phi$ and its complex conjugate $\tilde{\Phi} = i \tau_2
\Phi^{\ast}$ are necessarily complex.  If $Q^i, u^i_R$ and $d^i_R$
denote, respectively, the left-handed quark doublet of the
 $i^{th}$ generation and the charge $2/3$ and charge $-1/3$ right-handed
quarks of this same generation, one can write these Yukawa interactions
as
\be
{\cal{L}}_{\rm Yukawa} = \Gamma^u_{ij} {\bar{Q}}^i_L u^j_R \Phi +
\Gamma^d_{ij} \bar{Q}_L^i d^j_R \tilde{\Phi} + h.c.
\ee
When $\Phi$ and $\tilde{\Phi}$ are replaced by their vacuum expectation
values, the above interactions will give rise to complex mass matrices
for the quarks.  To go to a {\bf physical basis} where the quarks have real
diagonal masses, one must perform a unitary transformation on the quark
fields which, in general, will involve different unitary matrices for
the left-handed and right-handed fields and different matrices for the
charge $2/3$ and charge $-1/3$ fields:
\be
u_{L, R} = U^u_{L,R} u_{L,R}~~;~~ d_{L,R} = U^d_{L,R} d_{L,R}~.
\ee
As a result of this basis change, the interaction of the gauge fields
with the quarks, which used to be family diagonal, now no longer are
necessarily so.  Neutral current interactions, since they  involve
always $(U^u_{L,R})^{\dagger} U^u_{L,R} = 1$ or $(U^d_{L,R})^{\dagger}
U^d_{L,R} = 1$,
continue to be diagonal.  However, for charged current interactions
what enters after the basis change is the unitary matrix
\be
V_{CKM} = (U^u_L)^{\dagger} U^d_L~~,
\ee
or its adjoint. This gives rise to family mixing .

In this physical basis all the CP violating phase information present in
the Yukawa couplings is transferred to the Cabibbo Kobayashi Maskawa
matrix $V_{CKM}$.  Because this matrix is unitary, for $N_g$ generations
of quarks $V_{CKM}$ is parameterized by $\frac{1}{2} N_g (N_g - 1)$ real
angles and $\frac{1}{2} N_g (N_g + 1)$ phases.  However, not all of
these phases are physical since one can absorb $(2N_g - 1)$ phases by
appropriate quark field redefinitions \footnote{An overall phase cannot
be redefined away if CP is not conserved.}.  Thus in $V_{CKM}$ there are
in total $\frac{1}{2} (N_g - 1)(N_g - 2)$ physical phases. As a result,
as I alluded
to earlier, in the physically relevant case of 3 generations of quarks
there is only one physical phase $\delta$ in $V_{CKM}$, which is
responsible for all CP violating phenomena.  This is the CKM paradigm
\footnote{In principle, there is an analogous matrix to $V_{CKM}$ in the
leptonic sector of the standard model.  However, if the neutrinos are
massless, one can absorb this matrix entirely by redefining once more
the neutrino fields, since they are mass degenerate.  Because CP
violation in the lepton sector is connected with neutrino mass
generation, I shall not discuss it further here.
I note only that, for analogous
reasons,
CP violating effects in the quark sector
in the CKM paradigm  will vanish in
the limit when the quarks become mass degenerate.}.

\subsection{ The Wolfenstein Parameterization of the CKM Matrix}

The $3 \times 3$ unitary matrix $V_{CKM}$ characterizing charged current
weak interactions can be specified in many equivalent forms.  It proves
convenient to adopt a standard parametrization \cite{PDG} which admits
a simple and useful approximate form \cite{Wolf}.  One writes
\begin{eqnarray}
V_{CKM} &=& \left[ \begin{array}{clc}
V_{ud} & V_{us} & V_{ub} \\
V_{cd} & V_{cs} & V_{cb} \\
V_{td} & V_{ts} & V_{tb} \end{array} \right]\nonumber \\
&=& \left[\begin{array}{clc}
c_1 c_3 & s_1 c_3 & s_3 e^{-i \delta} \\
-s_1 c_2 - c_1 s_2 s_3 e^{i \delta} & c_1 c_2 - s_1 s_2 s_3 e^{i \delta}
& s_2 c_3 \\
s_1 s_2 - c_1 c_2 s_3 e^{i \delta} & -c_1 s_2 - s_1 c_2 s_3 e^{i \delta}
& c_2 c_3 \end{array} \right]~~,
\end{eqnarray}
where
\be
c_i \equiv \cos \theta_i~~;~~ s_i \equiv \sin \theta_i~~.
\ee
The above is well approximated to $0 (\lambda^4)$ by writing for the
$s_i$ the hierarchical parametrization \cite{Wolf}
\be
\sin \theta_1 = \lambda~~;~~ \sin \theta_2 = A \lambda^2~~;~~ \sin
\theta_3 = A \sigma \lambda^3~~.
\ee
Here $\lambda$ is essentially the sine of the Cabibbo angle and one has,
experimentally,
\be
\lambda \simeq \sin \Theta_c = 0.22~~,
\ee
while A and $\sigma$, as we shall see, turn out to be of $0(1)$.
In terms of the above, (Wolfenstein) parametrization one can write
$V_{CKM}$ to $0(\lambda^4)$ as
\bea V_{CKM} = \left[\begin{array}{clc}
1 - {\lambda^2\over 2} & \lambda & A \lambda^3 \sigma e^{-i \delta}
 \\
- \lambda & 1 - {\lambda^2\over 2} & A \lambda^2\\
A \lambda^3 (1 - \sigma  e^{i \delta}) & - A \lambda^2 & 1
\end{array} \right]~~,
\eea
or, more conventionally, writing $\sigma e^{-i \delta} = \rho - i \eta$
\bea
V_{CKM} = \left[\begin{array}{clc}
1 - {\lambda^2\over 2} & \lambda & A \lambda^3 (\rho - i \eta)
 \\
- \lambda & 1 - {\lambda^2\over 2} & A \lambda^2  \\
A \lambda^3 (1 - \rho - i \eta) & - A \lambda^2 & 1
\end{array} \right]~~.
\eea

I note for future use that to $0 (\lambda^4)$ the CKM matrix has only
two elements which have an imaginary part, $V_{ub}$ and $V_{td}$.
Furthermore, from the form of the matrix one sees that information on
the, still to be determined, parameters $A$ and $\sigma$ (or
$\sqrt{\rho^2 + \eta^2)}$ necessitates measurements involving $b$ quarks,
with $A$ being fixed by $V_{cb}$ and $\sigma$ (or $\sqrt{\rho^2 +
\eta^2)}$ by the ratio of $\vert V_{ub} \vert/\vert V_{cb} \vert$.
Obviously, information on the phase $\delta$ (or the CP violating
parameter $\eta$) can be gotten from the measurements of CP violating
phenomena in the $K^0 - {\bar{K}}^0$ complex.  However, because $\delta$
(and $\eta$) also enter in $V_{td}$ some useful information on this
parameter can also be garnered from non CP violating phenomena, like
$B^0_d - {\bar{B}}^0_d$ mixing which depend on this matrix element.  We
will return to discuss how well $A$, $\sigma$ and $\delta$ (or $A, \rho$
and $\eta$) are determined at present experimentally, after we discuss
the predictions of the CKM paradigm for CP violation in the Kaon
system.

\section{CP Violation in the Kaon System}

To compare the experimental values of the various CP violating
parameters which are measured in the $K^0 - {\bar{K}}^0$ complex and
which we discussed in Section II, it is necessary to develop a bit of
formalism.  This formalism will also be relevant later on when I will
discuss CP violation in the $B$ system.

\subsection{Two State Formalism}

Neutral particle - antiparticle systems $( P - \bar{P}$ systems), like
those formed by a $K^0 \sim d \bar{s}$ and a ${\bar{K}}^0 \sim \bar{d}
s$ or by a $B^0_d \sim d \bar{b}$ and a ${\bar{B}}^0_d \sim \bar{d} b$
meson, provide very nice examples of quantum mechanics at work.  The
individual states in these systems, $P$ and $\bar{P}$, are unstable due to
the weak interactions ($\Delta P = \pm 1$ processes).  Furthermore, $P$
and $\bar{P}$ can mix with each other via a 2nd order weak process
$(\Delta P = \pm 2$ processes).  It is useful to describe the decay and
the mixing in the $P - \bar{P}$ complex by means of an effective $2
\times 2$ Hamiltonian \cite{LW}

\be H_{\rm eff} = M - {i\over 2} \Gamma~~,
\ee
characterized by Hermitean mass, $M$, and decay, $\Gamma$, matrices. The
time evolution of the system is then described by the 2-state
Schroedinger equation:
\be
i \frac{\partial}{\partial t} \left(\begin{array}{c}
P\\ {\bar{P}}\end{array} \right) = H_{\rm eff}~~\left(\begin{array}{c}
P\\ {\bar{P}} \end{array} \right)~~.
\ee
If CPT is conserved, as I shall assume in what follows, then the
matrices $M$ and $\Gamma$ have a further constraint on them besides
their Hermiticity, namely
\be
M_{11} = M_{22}~~;~~ \Gamma_{11} = \Gamma_{22}~.
\ee
This constraint just reflects the simple fact that CPT requires
particles and antiparticles to have the same mass and the same lifetime.
 If in addition CP were to be a good symmetry - in the phase convention
where $CP \vert P> = \vert \bar{P} >$ - one would have the further
restriction that
\be
M_{12} = M^{\ast}_{12}~~;~~ \Gamma_{12} = \Gamma^{\ast}_{12}~~.
\ee
That is, the matrices $M$ and $\Gamma$ would be real.  Obviously, in the
CKM paradigm this will not be so because of the presence of the phase
$\delta$.

It is straightforward to deduce the physical eigenstates of $H_{\rm eff}$.
These are the states $\vert P_{\pm} >$ which have the simple time
evolution
\be
\vert P_{\pm} (t) > = e^{- i m_{\pm} t} e^{- \frac{1}{2} \Gamma_{\pm} t}
\vert P_{\pm} >~~.
\ee
That is, they are definite mass eigenstates, $m_{\pm}$, which decay with
fixed rates $\Gamma_{\pm}$.  Assuming CPT conservation, but not assuming
CP conservation, one finds that the states
 $\vert P_{\pm} > $ are the following combinations of $\vert P> $ and
$\vert {\bar{P}} >$ states
\be \vert P_{\pm} > = \frac{1}{\sqrt{2 (1 + \vert \epsilon_P \vert^2)}}
\left\{ (1 + \epsilon_P) \vert P > \pm~ (1 - \epsilon_P) \vert \bar{P} >
\right\}~~.
\ee
Here the complex parameter $\epsilon_P$ characterizes the amount of CP
violation in the evolution of the system.  Obviously, if $\epsilon_P$
where to vanish then the physical states $\vert P_{\pm} >$ would be CP
eigenstates.  One finds that
\be
\frac{1 - \epsilon_P}{1 + \epsilon_P} = \frac{[ M_{12}^{\ast} -
\frac{i}{2} \Gamma^{\ast}_{12} ]^{1/2}}{[M_{12} - \frac{i}{2}
\Gamma_{12}]^{1/2}} \equiv \eta_P e^{i \Phi_p}~~.
\ee

Because the states $\vert P_{\pm} >$ are the ones that have a definite
time evolution, and these are superpositions of $\vert P >$ and $\vert
\bar{P} >$, it follows that if one produces at $t = 0$ a state
$\vert P >$ this state will evolve in time into a superposition of
$\vert P >$ and $\vert \bar{P} >$ states.  A simple calculation gives the
following formula for the resulting state - which I'll call $\vert
P_{\rm phys} (t) >$ - at time $t$:
\be
\vert P_{\rm phys} (t) > = e^{- \frac{i}{2} (m_+ + m_-) t}
e^{ -
\frac{1}{4} (\Gamma_+ + \Gamma_-) t} \{ a(t) \vert P > + b (t) \vert
{\bar{P}} > \}
\ee
where
\be
a(t) = \cos \frac{\Delta H t}{2}~~;~~ b(t) = i \eta_P e^{i \Phi_p}
\sin \frac{\Delta Ht}{2}~~.
\ee
Here the parameter $\Delta H$ contains information of the physical
parameters of the $ P - \bar{P}$ complex
\be
\Delta H = (m_+ - m_-) - \frac{i}{2} (\Gamma_+ -  \Gamma_-)~~.
\ee
The parameter $b (t)$ above contains information on CP violation in the
system.  However, in general, this will not be the only place where CP
violating effects can enter.

Besides CP violation arising through the evolution and
mixing of $P$ and $\bar{P}$, there can be CP violating phases which enter
directly in the decay amplitudes of $P$ and $\bar{P}$ to some final
states $f$.  In the CKM paradigm, since $V_{CKM}$ enters precisely in
$\Delta P = \pm 1$ decays, one expects to have non trivial CP violating
phases in the decay amplitudes $A (P \to f)$.  Thus observable effects
of CP violation in the $P - \bar{P}$ complex are generally mixtures of
decay $(\Delta P = \pm 1)$ and mixing $(\Delta P = \pm2)$ CP violating
parameters.  Although in the CKM model all these parameters are related
to the phase $\delta$, it is difficult in general to relate directly CP
violating
observables to the underlying theory.  Nevertheless, we will see that
the CKM model has considerably different predictions for CP violation
in the Kaon complex than it does
in the $D^0$, $B^0_d$ and $B^0_s$ systems.  Kaons
are a special case since $\Gamma_+ \gg \Gamma_-$ and already specific
decays, like $K_L \to \pi^+ \pi^-$, are direct signals of CP violation.
In contrast, for the most part in Kaon CP violation one is
only able to give qualitative tests of
the theory.  On the other hand,
in the $B^0_d$ and $B^0_s$ systems, as
we shall see,
one
can disentangle better some of the dynamical complications which ensue
in trying to compare theory with experiment.  As a result,
one can
hope that future measurements of CP violation in these systems may
really provide quantitative tests of the theory.

\subsection{Neutral Kaon Amplitudes:~\protect $\epsilon$ and
\protect$\epsilon^{\prime}$}

The parameters $\epsilon$ and $\epsilon^{\prime}$, which we defined in
Sec. II, can be related in a straightforward manner to the $K$ decay
amplitudes and the mixing parameters $\epsilon_K$.  Working to lowest
order in small quantities one has
\bea
\vert K_S > &\simeq \frac{1}{\sqrt{2}} \{ (1 + \epsilon_K ) \vert K^0 > + (1 -
\epsilon_K) \vert {\bar{K}}^0> \} \nonumber \\
\vert K_L > &\simeq \frac{1}{\sqrt{2}} \{ (1 + \epsilon_K ) \vert K^0 > - (1 -
\epsilon_K) \vert {\bar{K}}^0> \} ~~.
\eea
Let me denote the amplitudes for a $K^0$ to decay into a $\pi \pi$ state
of isospin I by
\be < 2 \pi; I \vert T \vert K^0 > = A_I e^{i \delta_I}~~,
\ee
where $\delta_I$ is the $\pi - \pi$ scattering phase shift in the
channel of isospin $I$.  Then the corresponding amplitude for ${\bar{K}}^0$
decay is
\be
< 2 \pi; I \vert T \vert {\bar{K}}^0 > = A_I^{\ast} e^{i \delta_I}~~.
\ee
That is, this amplitude will have the same strong scattering
phase factor but, if there is
$\Delta S = 1$ CP violation, this amplitude will differ from that for
$K^0$ decay since it involves $A_I^*$.  This is what one expects
in the CKM paradigm.

Using the isospin decomposition
\bea
\vert \pi^+ \pi^- > &=& {\sqrt{\frac{2}{3}}}~ \vert~ 2 \pi; 0 > +
{\sqrt{\frac{1}{3}}}~ \vert~ 2 \pi; 2 > \nonumber \\
\vert \pi^0 \pi^0 >&=& {\sqrt{\frac{1}{3}}}~ \vert~ 2 \pi; 0 >  -
{\sqrt{\frac{2}{3}}}~ \vert~ 2 \pi; 2 > ~~,
\eea
and expanding in small quantities again, it is easy to derive formulas
for
\be
\eta_{+ -} = \frac{< \pi^+ \pi^- \vert T \vert K_L >}{< \pi^+ \pi^- \vert
T \vert K_S >} ~~;~~ \eta_{00} = \frac{< \pi^0 \pi^0 \vert T \vert
K_L >}{< \pi^0 \pi^0 \vert T \vert K_S >}
\ee
and whence formulas for $\epsilon$ and $\epsilon^{\prime}$.  Because
experimentally $\vert A_2 \vert / \vert A_0 \vert$ is small, of $0
(1/20)$, it suffices also to retain only terms in first order in this
quantity - the so called $\Delta I = 3/2$ suppression.  One finds
\bea
\epsilon &\simeq& \epsilon_K + i \frac{Im A_0}{Re A_0}\nonumber \\
\epsilon^{\prime} &\simeq& \frac{i}{\sqrt{2}} e^{i (\delta_2 - \delta_0)}
\frac{Re A_2}{Re A_0} \left[ \frac{Im A_2}{Re A_2} - \frac{Im
A_0}{Re A_0} \right]
\eea
The parameter $\epsilon^{\prime}$ which measures the difference between
$\eta_{+-}$ and $\eta_{00}$ is obviously suppressed by the small factor
of $Re A_2/Re A_0$, but this by itself cannot account for the very small
value of $\epsilon^{\prime}/\epsilon$ seen experimentally.

In the older literature, one often used a phase convention suggested by
Wu and Yang \cite{WY} to simplify the formulas for $\epsilon$ and
$\epsilon^{\prime}$.  Wu and Yang  made use of the freedom of choosing a
CP phase for how $\vert K^0 >$ and $\vert {\bar{K}}^0 >$ are related
under CP,

\be
CP \vert K^0 > = e^{i \xi} \vert {\bar{K}}^0 >~,
\ee
and chose $\xi$ so as to make $Im A_0 = 0$.  Then $\epsilon \equiv
\epsilon_K$.  I find it more physical to use the {\bf quark phase
convention} where $\xi = 0$ and the phases for $A_0$ and $A_2$ just
follow from the CKM phase that enters at the quark vertices.
Nevertheless, even in this case, one can still arrive at some
simplifications by using the fact that the $2 \pi$ intermediate state is
by far the dominant contribution in the width matrix $\Gamma$.

Using the formulas given previously, it is easy to show that the mixing
parameter $\epsilon$ is given by the equation
\be
\epsilon_K = \left[  \frac{ - Im M_{12} + \frac{i}{2} Im \Gamma_{12}}{i
\Delta H} \right]_K~~.
\ee
The kinematical parameter $(\Delta H)_K$ is related to the $K_L - K_S$
mass difference $\Delta m$ and the superweak phase $\phi_{sw}$
\be
i
(\Delta H)_K = \frac{1}{2} ( \Gamma_S - \Gamma_L) - i (m_L - m_S) \simeq
\sqrt{2} \Delta m e^{- i \phi_{sw}}~~.
\ee
Because of
the $2 \pi$ dominance
in the $K^0$ amplitudes one has that
\be
\Gamma_S - \Gamma_L \simeq \Gamma_S \sim 2 (Re A_0)^2~~,
\ee
while
\be
\Gamma_{12} \sim (A^{\ast}_0 )^2~~.
\ee
Thus
\be
\frac{Im~ \Gamma_{12}}{\Gamma_S - \Gamma_L} \simeq - \frac{Im A_0}{Re
A_0}~~.
\ee
Using this result, along with the definition of $\epsilon_K$ and the
fact that $\Delta m \simeq \frac{1}{2} (\Gamma_S - \Gamma_L)$, one easily
deduces that \cite{Cronin}
\be
\epsilon = \epsilon_K + i \frac{Im A_0}{Re A_0} \simeq
\frac{1}{\sqrt{2}} e^{i \phi_{sw}} \left[ - \frac{Im M_{12}}{\Delta m}
+ \frac{Im A_0}{Re A_0} \right]~~.
\ee

Several remarks are in order:
\begin{description}
\item{i.)} From the above formula for $\epsilon$ one sees that one
expects that the phase of $\epsilon$ should be the superweak phase
$\phi_{sw}$.  Since $\epsilon^{\prime} \ll \epsilon$, the phase of
$\epsilon$ is just that of $\eta_{+-}$ and indeed experimentally
$\phi_{+-} \simeq \phi_{sw}$.  This result is actually a CPT test, for
one can show that the inclusion of CPT violating effects, both in the
evolution of the $K^0 - {\bar{K}}^0$ system and in the decay amplitudes,
produces contributions to $\epsilon$ with a phase $\phi_{sw} + \pi/2$
\cite{Cronin}\protect\cite{BCDPQ}.

\item{ii.)} Since there are no CP violating phases in the semileptonic
decay amplitudes of $K^0$ and ${\bar{K}}^0$, the semileptonic asymmetry
$A_{K_{L}}$ can only depend on the mixing CP violating parameter
$\epsilon_K$.  A simple calculation then secures the formula
\be
A_{K_{L}} = 2 Re~ \epsilon_K~~.
\ee
Since $Re~ \epsilon_K = Re ~\epsilon \simeq Re~ \eta_{+-}$, one sees that
the observed relationship $A_{K_{L}} \simeq 2 Re~ \eta_{+-}$ follows
directly.  Again, the extent by which  $Re~ \epsilon -
\frac{1}{2} A_{K_{L}} $ differes from zero is a test of CPT, for if CPT
were not conserved both the formula for $\epsilon$ and $A_{K_{L}}$ would
contain further terms.  Using the experimental values for $A_{K_{L }},
\eta_{+-}$ and the phase $\phi_{+-}$ one finds \cite{RDP}
\bea
Re~ \epsilon - \frac{1}{2} A_{K_{L}} =  \left\{
\begin{array}{cc} ( - 0.6 \pm 0.7 ) \times 10^{-4} & \mbox {\protect
\cite{PDG}} \\
(0.3 \pm 0.7 )\times 10^{-4} & \mbox{\protect \cite{EZ31}}
\end{array}\right.~~,
\eea
where the two different values above result from using, respectively, the
PDG value for $\phi_{+-}$ and that of E731.
\item{iii.)} The values for the $\pi \pi$ phase shift difference
\cite{GM}, $\delta_2 - \delta_0 = - (45 \pm 6)^0$, is such that the
phase of $\epsilon^{\prime}$ is also very near $45^0$.  Thus to a very
good approximation
\be
Re \frac{\epsilon^{\prime}}{\epsilon} \simeq
\frac{\epsilon^{\prime}}{\epsilon} \simeq
\left[
\frac{{\rm Re} A_2}{{\rm Re} A_0}\right]
\left[\frac{ \bgfrac{ {\rm Im} A_2}{ {\rm Re} A_2} -
\bgfrac{ {\rm Im} A_0}{ {\rm Re} A_0}
}{ -\bgfrac{ {\rm Im}M_{12}}{\Delta m}
+ \bgfrac{ {\rm Im} A_0}{ {\rm Re} A_0}
} \right]~~.
\ee
Experimentally $\epsilon^{\prime}/\epsilon$ is much
below the $\Delta I
= 3/2$ suppression factor of $Re A_2/Re A_0 \sim 1/20$.  Hence, barring
an accidental cancellation in the numerator one deduces that
\be
\vert \frac{Im A_0}{Re A_0}|  \ll \vert \frac{Im M_{12}}{\Delta m}|
\ee
and hence that $\vert \epsilon \vert$ should be well
approximated by the
simple formula
\be
\vert \epsilon \vert \simeq \frac{\vert Im M_{12} \vert}{\sqrt{2} \Delta
m}~~.
\ee
That is, $\vert \epsilon \vert$ is essentially
only a result of mixing CP
violation.
\end{description}

\subsection{ Boxes and Penguins - Standard Model
Predictions.}

In the standard electroweak model $M_{12}$ arises from the 2nd order
weak box diagram shown in Fig. 1.  This diagram gives rise to an
effective $\Delta S = 2$ Lagrangian
\be
{\cal{L}}_{\Delta S = 2} = A_{\rm box} [ \bar{d} ( \gamma_{\mu} (1 - \gamma_5)
s) (\bar{d} \gamma^{\mu} (1 - \gamma_5) s) ] + h.c.,
\ee
where the coefficient $A_{\rm box}$ is easily read out from Fig. 1. One has

\begin{figure}
\vspace*{6cm}
\caption[ ]{Box diagram giving rise to the mass mixing matrix element
\protect$M_{12}$.}
\end{figure}

\bea
A_{\rm box} &=& \int \frac{d^4 q}{(2 \pi)^4} D^W_{\mu \alpha} (q)
D^W_{\nu_{\beta}} (q)~~. \nonumber \\
&& \cdot \sum_{ij} \lambda_i [\gamma^{\mu} (1 - \gamma_5) D_i
(q) \gamma^{\nu} ( 1 - \gamma_5) ] \nonumber \\
&& \lambda_j [ \gamma^{\beta}( 1 - \gamma_5) D_j
(q) \gamma^{\alpha} (1 - \gamma_5)]
\eea
Here $D^W_{\mu \alpha} (q)$ and $D_i (q)$ are the $W$ and $i^{th}$ fermion
propagators, respectively, while the coefficients $\lambda_i$ involve
the following products of CKM matrix elements:
\be
\lambda_i = V_{i s} V_{id}^{\ast}~~.
\ee

The integral in $A_{\rm box}$ is potentially quadratically divergent.
However, the unitarity of the CKM matrix implies that
\be
\lambda_u + \lambda_c + \lambda_t = 0~~.
\ee
Eliminating $\lambda_u$ in the expression for $A_{\rm box}$ gives
differences of terms and these lead to a convergent expression for
$A_{\rm box}$.  This is the celebrated GIM mechanism \cite{GIM}.  A
straightforward calculation \cite{IL} secures the following formula for
$A_{\rm box}$.
\be
A_{\rm box} = \frac{G_F^2}{16 \pi^2} \{
 \lambda_c^2 m_c^2 \eta_1 + \lambda_t^2 m_t^2 f_2 (y_t)
\eta_2
+ 2 \lambda_c \lambda_t m_c^2 ( \ell n \frac{m_t^2}{m_c^2} + f_3 (y_t)
) \eta_3 \}~~.
\ee
Here $f_2 (y_t)$ and $f_3(y_t)$ are kinematical functions which are
weakly dependent on the ratio of the top mass to the $W$-mass, $y_t =
m_t^2/M_W^2$.  One finds \cite{IL}
\bea f_2 (y_t) &=& 1 - \frac{3y_t (1 +
y_t)}{4 (1 - y_t)^2} \left[ 1 + \frac{2 y_t}{1 - y_t^2} \ell n y_t
\right] \nonumber \\
f_3 (y_t) &=& \frac{ - 3y_t}{4 (1 - y_t)} \left[ 1 + \frac{y_t}{1 - y_t}
\ell n y_t \right]~~.
\eea
The coefficient $\eta_i$ are $QCD$ short distance corrections to the box
graph of Fig. 1, arising from festooning this graph with gluons.  These
coefficients have been calculated by Gilman and Wise \cite{GW} and they
find, approximately,
\be
\eta_1 \simeq 0.7~~;~~ \eta_2 \simeq 0.6~~;~~ \eta_3 \simeq 0.4~~.
\ee

To be more precise, the actual values for the
$\eta_i$ are dependent on the
scale $\mu$ at which the operator entering in ${\cal{L}}_{\Delta S =
2}$ is normalized.  However, the matrix element of this operator also
depends on $\mu$, with physical cogency requiring that the product of
the $\eta_i$ and the matrix element be $\mu -$ independent.  It has
become conventional to write this matrix element in terms of a constant
$B_K (\mu)$, which is normalized so that the value $B_K = 1$
corresponds to the vacuum insertion approximation to this matrix
element.  That is
\be
< K^0 \vert \bar{d} \gamma_{\mu} (1 - \gamma_5) s \bar{d} \gamma^{\mu} (
1 - \gamma_5) s \vert {\bar{K}}^0 > = \frac{8}{3} f^2_K M^2_K B_K
(\mu^2)~~,
\ee
with $f_K \simeq 160~ MeV$ being the Kaon decay constant. The best
evaluation of $B_K (\mu)$ obtained in lattice $QCD$, for the $\mu$
values which give the $\eta_i$ values given above, is \cite{Martinelli}
\be
B_K (\mu) = 0.80 \pm 0.20~~.
\ee

For the computation of $\epsilon$ one needs to know $Im M_{12}$ and one
has
\be
Im M_{12} = \frac{Im < K^0 \vert {\cal{L}}_{\Delta S = 2} \vert {\bar{K}}^0
>}{2 M_K} =  \frac{4}{3} f_K^2 M_K B_K Im A_{\rm box}~~.
\ee
$A_{\rm box}$ only contains an imaginary part as a result of the
nontrivial phase $\delta$ in the CKM matrix.  To the accuracy one is
working here, it does not suffice to approximate $V_{cd}$ to
$0(\lambda^3)$ but one must retain its contributions to $0
(\lambda^5)$,
where one has
\be
V_{cd} = - \lambda [ 1 + A^2 \sigma \lambda^4 e^{ + i \delta} ]~~.
\ee
To leading order in $\lambda$ in both the real and imaginary parts one
finds
\bea
\lambda^2_c &=& (V_{cs} V^{\ast}_{cd} )^2 \simeq \lambda^2 - 2i A^2 \sigma
\lambda^6 \sin \delta \nonumber \\
\lambda_t^2 &=& (V_{ts} V_{td}^{\ast} )^2 \simeq A^4 \lambda^{10} \left\{ [(1
- \sigma \cos \delta)^2 - \sigma^2 \sin^2 \delta ] + 2 i \sigma ( 1 -
\sigma \cos \delta ) \sin \delta \right\} \nonumber \\
2 \lambda_c \lambda_t &=& 2 (V_{cs} V_{cd}^{\ast} V_{ts} V_{td}^{\ast} )
= 2 A^2 \lambda^6 \left\{ (1 - \sigma \cos \delta) + i \sigma \sin
\delta \right\}~~.
\eea
Even though $m^2_t \gg m_c^2$, for the real part of $A_{\rm box}$ the
only relevant piece is that proportional to $\lambda_c^2$ and
$\lambda_c^2 \simeq \lambda^2$.  However, for $Im A_{\rm box}$ the
contributions of the $\lambda_c^2$ and $2 \lambda_c \lambda_t$ terms are
comparable, being both proportional to $\lambda^6$.  Furthermore, even
though $Im \lambda_t^2 \sim \lambda^{10}$, the multiplying factor of
$m_t^2$ rather than $m_c^2$ in $A_{\rm box}$ does not allow one to
neglect this term.

Collecting all this information together, one arrives at the following
master formula for $\epsilon$ in the standard model \cite{BSS}:
\bea
\vert \epsilon \vert \simeq \frac{\vert Im M_{12}\vert}{\sqrt{2} \Delta
m} &\simeq& \frac{G_F^2 f_K^2 M_K}{6 \sqrt{2} \pi^2 \Delta m} B_K
[A^2 \sigma \lambda^6 \sin \delta ]~~. \nonumber \\
&&\cdot \biggl\{ m_c^2 [ - \eta_1 + \eta_3 \left( \ell n \frac{m_t^2}{m_c^2} +
f_3 (y_t) \right) ] + \nonumber \\
&&~~~~~~~~~+ m_t^2 \eta_2 f_2 (y_t)\biggl[A^2 \lambda^4 (1 - \sigma \cos
\delta)\biggr] \biggr\}~~.
\eea
I will discuss shortly a more detailed comparion of this formula with
experiment.  However, it is useful to get first an order of magnitude
estimate of the expected size of $\epsilon$.  The $K_S - K_L$ mass
difference $\Delta m$, neglecting possible long distance
contributions\cite{Wolfenstein}, is given by
\be
\Delta m \simeq Re  \frac{< K^0 \vert {\cal{L}}_{\Delta S = 2} \vert
{\bar{K}}^0>}{2m_K}~~.
\ee
Thus a rough estimate of $\epsilon$ is provided by

\begin{eqnarray}
\vert \epsilon \vert \simeq \frac{Im A_{\rm box}}{Re A_{\rm box}}
&\simeq& A^2 \sigma \lambda^4 \sin \delta \biggl\{\bigl[ - \eta_1 + \eta_3
\bigl(\ln \frac{m^2_t}{m_c^2} + f_3 (y_t)\bigr)\bigr] \nonumber \\
&+& \frac{m_t^2}{m_c^2} \eta_2 f_2 (y_t) \lambda^4 A^2 ( 1 - \sigma \cos
\delta) \biggr\}
\end{eqnarray}
Since the quantity in the curly bracket is of $0(1)$ - and so as we
shall see are $A$ and $\sigma$ - one sees that in the standard model
$\epsilon$ is of order
\be
\epsilon \sim \lambda^4 \sin \delta~~.
\ee
Because $\lambda^4 \sim 2 \times 10^{-3}$ one sees that in the standard
model $\epsilon$, and therefore CP  violation in the Kaon system, is
small {\bf not} because the phase $\delta$ is particularly small, but
because the interfamily mixing (represented by the factor of
$\lambda^4$) is small.

Dressing of the box graph of Fig. 1 by gluons gives the QCD corrections
to $\vert \epsilon \vert$ characterized by the $\eta_i$ coefficients
entering in Eq. (59).  For $\epsilon^{\prime}$, however, gluonic effects
are {\bf fundamental}, for this quantity vanishes in the limit that
$\alpha_s \to 0$.  The relevant diagrams that contribute to
$\epsilon^{\prime}$ are the, so called, Penguin diagrams of Fig. 2
\cite{SVZ}\cite{GWCP}.
\begin{figure}[h]
\vspace*{6cm}
\caption[]{Gluonic Penguin diagram contributing to
\protect$\epsilon^{\prime}.$}
\end{figure}
The calculation of these diagrams is made simple by noting the following
\cite{PTAS}:

\begin{description}
\item{i)} Although each individual diagram, containing an $u, c$ or $t$
propagator, is divergent, the piece that is relevant for
$\epsilon^{\prime}$ is convergent since it is the part of the Penguin
amplitude which is proportional to $q^2$ - the gluon momentum transfer.
\item{ii)} This $q^2$ factor with the $1/q^2$ factor from the gluon
propagator, leads to an effective 4-Fermi interaction.

\item{iii)} The leading contribution for the Penguin diagrams is easily
computed in a $4$-Fermi limit for the $W$ exchange, being simply
proportional to the logarithmic divergent piece of the diagrams in this
limit.  To get
the physical relevant answer, one then only needs to
replace $\ell n \Lambda^2$ by $\ell n M_W^2$, with a $\Lambda$ being the
cutoff.
\end{description}

Using the above, it is straightforward to derive the effective Penguin
interaction for each quark $i$ by computing the logarithmic divergent
piece of the (subtracted) graph shown in Fig. 3.
\begin{figure}
\vspace*{6cm}
\caption[]{Effective (subtracted) graph needed for Penguin computation.}
\end{figure}
One finds in this way \cite{GWCP}
\be
{\cal{L}}_{\rm gluonic Penguin} = \frac{G_F}{\sqrt{2}}~
\frac{\alpha_s}{12 \pi} A_{P} \left\{ (\bar{s} \gamma_{\mu} (1 -
\gamma_5) \lambda_a d) \cdot \sum_q (\bar{q} \gamma^{\mu} \lambda_a q)
\right\}
\ee
where $\lambda_a$ are $SU(3)$ matrices.
The coefficient $A_{P}$ in the limit in which $m_i \ll M_W $ -
something we know is not true for $m_t$, but which will be corrected
below - is given simply by
\be
A_{P} = \sum_i \lambda_i \ell n~ M^2_W/m_i^2~~.
\ee
Using the CKM unitarity [Eq. (58)] this can be rewritten as
\be
A_{P} = \lambda_t~ \ell n~ m_c^2/m_t^2 + \lambda_u~ \ell n~ m_c^2 /m_u^2~~.
\ee
For CP violating phenomena only the imaginary part of the above is
relevant, and since $Im \lambda_u = 0$, one sees that what effectively
dominates is the $t$-quark diagram in Fig. 2 with
\be
Im \lambda_t = - A^2 \lambda^5 \sigma \sin \delta~~.
\ee

Because all quark species $q$ in Eq. (71) are summed over
 with equal weight,
it is clear that the gluonic Penguin operator carries $I = \frac{1}{2}$.
 Thus, it contributes only to $Im A_0$ in the formula [ c.f. Eq. (42)]
for $\epsilon^{\prime}$.  One has
\be Im A_0 = C_P < \pi \pi; 0 \vert
\bar{s} \gamma^{\mu}(1 - \gamma_5) \lambda_a d
\sum_q (\bar{q} \gamma_{\mu} \lambda_a q) \vert K^0>
\ee
where
\be
C_P = [ \frac{G_F}{\sqrt{2}} \lambda ] \cdot [ \frac{\alpha_s}{12 \pi}
{}~\ell n \frac{m_t^2}{m_c^2} ] \cdot [ A^2 \lambda^4 \sigma \sin \delta ]~~.
\ee
The three terms in square brackets above characterize different physical
contributions to $\epsilon^{\prime}$.  $G_F \lambda/\sqrt{2}$
is the strength associated with a typical Kaon weak decay matrix
element.  Indeed $Re A_0$ has precisely this strength:
\be
Re A_0 = \frac{G_F \lambda}{\sqrt{2}} < \pi \pi; 0 \vert \bar{s}
\gamma^{\mu} ( 1 - \gamma_5) u \bar{u} \gamma_{\mu} (1 - \gamma_5) d
\vert K^0>~~.
\ee
The factor of $ \frac{\alpha_s}{12 \pi} ~\ell n~ \frac{m_t^2}{m_c^2}$
reflects the fact that this contribution to $\epsilon^{\prime}$ arises
as a result of $QCD$ and would vanish in the limit of degenerate quark
masses.  Finally the last bracket in Eq. (76) contains the same family
mixing suppression factor $A^2 ~\lambda^4 \sigma \sin \delta$ that enters in
$\epsilon$, which is prototypical of the CKM paradigm.

The coefficient $C_P$ in Eq. (76) is not quite correct, since it was
derived in the limit that $m_t \ll M_W$.  This can be readily remedied
\cite{AF}.  Furthermore one needs also to incorporate higher order QCD
corrections \cite{GP} into $C_P$.
The net result of doing both these things
is to replace the
second factor in Eq. (76) by a more complicated function than that given
in this equation.  For a typical range of $m_t$ values $(100~ GeV < m_t <
200~GeV)$ and for a QCD
scale $\mu$ appropriate to the problem at hand $(\mu
\sim m_K)$, this more accurate calculation gives for this factor a
numerical value of about $0.1$~\cite{AF}:
\be
\frac{\alpha_s}{12 \pi}~ \ell n \frac{m_t^2}{m_c^2} \to  0.085 \pm
0.035~,
\ee
where the error in the above
includes that produced by variations in the QCD scale
and on the value of $m_t$.

Besides this factor,  the contribution of $\epsilon^{\prime}$
relative to $\epsilon$ is further reduced by the $\Delta I = 3/2$
suppression factor contained in the ratio
\be
\frac{1}{\sqrt{2}} \frac{Re A_2}{Re A_0} \simeq 0.032~~,
\ee
where the numerical value follows from the experimental measurement of
the rate for $K^+ \to \pi^+ \pi^0$ relative to that of $K_S \to \pi^+
\pi^-$.  Hence, as an order of magnitude estimate for
$\epsilon^{\prime}/\epsilon$,  one has
\be
\epsilon^{\prime}/\epsilon \sim \frac{1}{\sqrt{2}} \frac{Re A_2}{Re A_0}
[\frac{\alpha_s}{12 \pi} ~\ell n \frac{m_t^2}{m_c^2}]_{\rm eff}~
\simeq 3 \times 10^{-3}~~,
\ee
which is in rough accord with the experimental result given in Eq. (4).

The above estimate ignores the fact that $\epsilon^{\prime}$ and
$\epsilon$ involve quite different operator matrix elements.
Nevertheless, it is quite gratifying to see that in the CKM paradigm
$\epsilon^{\prime}/\epsilon \ll 1$ as a result of the $\Delta I = 3/2$
suppression and the fact that $\epsilon^{\prime}$ vanishes as $\alpha_s
\to 0$.  In fact, in detail the situation regarding $\epsilon^{\prime}$ is
actually much more complicated.  In addition to the gluonic Penguin
diagrams of Fig. 2, there are electroweak Penguin diagrams - of which
some examples are
shown in
Fig. 4 - which contribute to the $\Delta I = 3/2$ amplitude $Im A_2$.
Although these electroweak Penguin contributions are suppressed relative
to those of the gluonic Penguins by a factor of
$\alpha/\alpha_s$, these
contributions are not negligible since they have no $\Delta I = 3/2$
suppression.
\begin{figure}
\vspace{6cm}
\caption[]{Some examples of electroweak Penguin diagrams contributing
to \protect$Im A_2$.}
\end{figure}
{}From Eq. (42) one sees that the term in $\epsilon^{\prime} $
proportional to $Im A_2$ is not divided by $Re A_0$ but by $Re A_2$.
Hence, really, electroweak Penguins are relatively enhanced.  In effect,
one has:
\be
\frac{Im A_2}{Re A_2} = \frac{Re A_0}{Re A_2} \cdot \frac{Im A_2}{Re
A_0} \simeq 20~ \frac{Im A_2}{Re A_0}~~.
\ee
Furthermore, as shown by Flynn and Randall \cite{FR}, these contributions
for large $m_t$ grow like $m_t^2$, not as $\ell n~m_t^2$, and tend to
cancel those of the gluonic Penguins.  Thus an accurate estimate of
$\epsilon^{\prime}/\epsilon$ in the CKM paradigm requires considerable
more care.  I will return to this point below.

\subsection{Comparison with Experiment}

In the CKM paradigm the experimental value for $\epsilon$ can be used to
determine the phase $\delta$.  Although $\epsilon \sim \sin
\delta$, it is not simple to extract a precise value of $\delta$ from
the precisely measured value of $\epsilon$.  This is because the
relationship between $\epsilon$ and $\sin \delta$ is a function of other
parameters, like $A, \sigma, m_t, m_c$ and $B_K$ which are
relatively poorly known.  In what follows I shall use the value of
$B_K$ given by the lattice QCD computation of this parameter [c.f.
Eq. (63)] and shall take $m_c = 1.4~GeV$, letting $m_t$ range.  For this
value of $m_c$ and $m_t < 200~GeV$, the formula for $\vert \epsilon \vert$, Eq.
(67), is well approximated by \cite{BSS}.\footnote{The error in Eq. (82)
is mostly due to the uncertainty in the hadronic matrix element,
typified by \protect$B_K.$}
\be
\vert \epsilon \vert = (2.7 \pm 0.7) \times 10^{-3} A^2 \sigma \sin
\delta \{ 1 + \frac{4}{3} (\frac{m_t}{M_W} )^{1.6} A^2 (1 - \sigma \cos
\delta) \}~~.
\ee
To proceed further one needs a value for the parameters $A$ and $\sigma$
which enter in the CKM matrix.

The parameter $A$ is essentially fixed by the $V_{cb}$ matrix element,
while $\sigma$ depends on how well one can determine the ratio of
$V_{ub}$ to $V_{cb}$ in $V_{CKM}$.
The values for $V_{cb}$ obtained from studying {\bf inclusive}
semileptonic decays of $B$ mesons to hadrons with charm $[ B \to X_c
\ell \nu_e]$ typically seem to be somewhat larger than those obtained by
analyzing specific {\bf exclusive} modes. In general, one expects that
the former analysis  be somewhat more reliable, as one can exercise
some control by demanding a simulaneous fit of the lepton spectrum
\cite{ACM}.  However, new theoretical ideas, connected with a heavy
quark expansion \cite{IW} can be used to give absolute predictions for
certain exclusive modes - like $B \to D^{\ast}~\ell \nu_e$ at zero
recoil \cite{Neubert}.  The values of $V_{cb}$ extracted by these means
potentially should be the most accurate, once there are enough
statistics for these processes.  In Table 1, I display a representative
set of results for $V_{cb}$ and $A$ obtained by these different
techniques.
\begin{table}
\caption{Representative Results for \protect$V_{cb}$ and \protect$A$}
\vspace*{.2cm}
\begin{center}
\begin{tabular}{| l | l | l |} \hline
{}~~~~Technique &~~~~$\vert V_{cb} \vert$ &~~~~$A$ \\ \hline
Inclusive Spectrum \protect\cite{Lus} & $0.047 \pm 0.004$ & $0.97 \pm
0.08$ \\ \hline
Exclusive Decay \protect\cite{Lus} & $0.041 \pm 0.006$ &
\protect$0.85 \pm 0.12$ \\ \hline
Heavy Quark Limit \protect\cite{Neubert} & $0.045 \pm 0.007 $ & $0.93
\pm 0.14$ \\ \hline
Heavy Quark Limit \protect\cite{Burd} & $0.041 \pm 0.005$ & $0.85 \pm
0.10$ \\ \hline
\end{tabular}
\end{center}
\end{table}
A sensible choice, and the one I shall adopt, appears to be to take
\be
\vert V_{cb} \vert = 0.043 \pm 0.005~~;~~ A = 0.90 \pm 0.10~~.
\ee

To extract $\vert V_{ub} \vert/\vert V_{cb} \vert$ from experiment one
studies the semileptonic decays of $B$ mesons $(B \to X_u \ell
\nu_{\ell})$
in a region of momentum of the emitted lepton $(p_{\ell} > 2.3~ GeV)$
which insures kinematically that the hadronic states $X_u$ do not
contain a charmed quark.  That is, for $p_{\ell} > 2.3~ GeV$ the data
should only measure decays in which the transition $b \to u$ occurred.
However, to extract a value of $\vert V_{ub}\vert$ from this analysis
is non trivial, since one must be able to estimate precisely the
hadronic matrix elements involved in the $B \to X_u$ transition.  When
one does this estimate by employing, as in the ACM model \cite{ACM}, a
parton model - which is sensible in my mind, since one is summing over
all states $X_u$ - one gets a fairly large value for this matrix element
and hence a rather small value for $\vert  V_{ub}\vert /\vert V_{cb}
\vert$.  On the other hand, if one estimates the transition $B \to X_u$
by summing only over some (assumed dominant) exclusive channels, as in
the ISGW model \cite{ISGW}, the strength of the transition is smaller
and, consequently, one deduces a larger value for $\vert V_{ub}
\vert/\vert V_{cb}\vert.$

Using only the more recent and more accurate data obtained by CLEO II,
Cassel \cite{Cassel} quotes the following values for $\vert V_{ub}
\vert /\vert V_{cb} \vert$ extracted, respectively, using the ACM model
\cite{ACM} and the ISGW model \cite{ISGW}.
\begin{figure}
\vspace{6cm}
\caption[]{Box graph which gives rise to~\protect$B^0_d - {\bar{B^0_d}}$
 mixing.}
\end{figure}

\bea
\bigg| \frac{V_{ub}}{V_{cb}} \bigg|& = 0.07 \pm 0.01 \leftrightarrow~ \sigma
= 0.32
\pm 0.06~~~{\rm ACM ~Model} \nonumber \\\nopagebreak
\bigg| \frac{V_{ub}}{V_{cb}} \bigg| &=  0.11 \pm 0.02 \leftrightarrow~ \sigma
= 0.50 \pm 0.09~~~{\rm ISGW~ Model}
\eea
As a preferred value, Cassel takes the average of these two results and
expands somewhat the errors by including other model uncertainties
\cite{Cassel}:
\be
\bigg| \frac{V_{ub}}{V_{cb}} \bigg| = 0.085 \pm 0.045 \leftrightarrow
\sigma = 0.39 \pm 0.21.
\ee

It has become conventional to present the result of a CKM analysis of
the data as contour plots in the $\rho - \eta$ plane, where recall $\rho
= \sigma \cos \delta$ and $\eta = \sigma \sin \delta$.  The theoretical
formula for $\vert \epsilon \vert$ given in Eq. (82), using the
experimental value for $\vert \epsilon \vert \simeq \vert
\eta_{+-} \vert =
2.27 \times 10^{-3}$ and the value
of $A$ from Eq. (83), gives the following
constraint on these parameters:
\be 1 = (0.96 \pm 0.33) \eta \{ 1 + (1.08 \pm
0.24)(\frac{m_t}{M_W})^{1.6} (1 - \rho)\}~~.
\ee
In addition,
since $\sigma = \sqrt{\rho^2 + \eta^2}$, Eq. (85) constrains $\rho$ and
$\eta$ to an annular region centered at $\rho = \eta = 0$:
\be
\sqrt{\rho^2 + \eta^2} = 0.39 \pm 0.21~~.
\ee
There is a third constraint on these parameters which comes from $B^0_d -
{\bar{B^0_d}}$ mixing.  The amount of mixing is governed by the mass
difference $(\Delta m)_{Bd}$ in this system, which in turn is fixed by
the box graphs shown in Fig. 5. These graphs are totally dominated
by the contribution in which the
fermions in the loop are top quarks. As a result, the amount of $B^0_d -
{\bar{B^0_d}}$ mixing gives a measure of
\be
\vert V_{td} \vert^2 = A^2 \lambda^6 [ (1 - \rho )^2 + \eta^2]~~.
\ee
Thus the constraint coming from the experimentally determined value of
$B^0_d - {\bar{B^0_d}}$ mixing is another annulus in the $\rho - \eta$ plane,
this time centered at the point $\rho = 1, \eta = 0$.

One can measure the amount of $B_d^0 - {\bar{B^0_d}}$ mixing by
determining the ratio of ``wrong" to ``right" sign leptons in the decay
of a $\vert (B_d)_{\rm phys}>$ state.  Recall from our treatment in Sec.
IV.A that this state was one which at $t = 0$ was a pure $\vert B_d > $ state
but which, because of the possibility of mixing, evolved in time into a
linear superposition of $\vert B_d>$ and $\vert {\bar{B_d}} >$ states [cf.
Eq. (34)].  Physically, only ${\bar{B}}_d$ states decay semileptonically
into {\bf negatively} charged leptons.  Thus the ratio
\be
\chi_d = \frac{\Gamma \Bigl((B_d)_{\rm phys} \to \ell^- \bar \nu_{\ell}
X\Bigr)}{\Gamma \Bigl((B_d)_{\rm phys} \to \ell^+ \nu_{\ell} X
\Bigr) +  \Gamma \Bigl((B_d)_{\rm phys} \to \ell^- \bar \nu_{\ell} X\Bigr) }
\ee
is a measure of $B^0_d - {\bar{B^0_d}}$ mixing.

The quantity $\chi_d$ is readily calculated using Eq. (34).  For the
$B^0_d - {\bar{B^0_d}}$ system there is some simplification since the
width difference $\Gamma_+ - \Gamma_-$ is small both compared to the
widths $\Gamma_+ $ and $\Gamma_-$ themselves and to the mass difference
$\Delta m = m_+ - m_-$\cite{Punk}.  Thus, writing $m_+ + m_- = 2
m_{B_{d}}$
and $\Gamma_+ + \Gamma_- = 2 \Gamma_{B_{d}}$ and using the above
approximations, Eq. (34) for the $\vert (B_d)_{\rm phys} >$ state
simplifies to:
\bea
\vert (B_d)_{\rm phys} (t) > = e^{-i m_{B_{d}} t}
e^{-\frac{1}{2} \Gamma_{B_{d}}t} \Bigl\{ \cos \frac{\Delta m_{B_{d}} t}{2}
\vert
 B_d >
+ \nonumber \\
+ i \eta_{B_{d}} e^{i \Phi_{B_{d}}} \sin \frac{\Delta m_{B_{d}}t}{2} \vert
{\bar{B_{d}}} > \Bigr\}~~.
\eea
A simple calculation then gives for $\chi_d$ the following formula
\be
\chi_d = \frac{x_d^2}{2 ( 1 + x_d^2)}
\ee
where
\be
x_d = (\frac{\Delta m}{\Gamma})_{B_{d}} = \tau_{B_{d}}(\Delta
m)_{B_{d}}~~.
\ee

The recent compilation of Cassel \cite{Cassel} gives the world average
value
\be
\chi_d = 0.145 \pm 0.018 \pm 0.018
\ee
or
\be
x_d = 0.64 \pm 0.06 \pm 0.06~~.
\ee
The above value for $x_d$, along with a value for the $B_d$ lifetime,
can be used in conjunction with the formula for $(\Delta m)_{B_{d}}$,
obtained by evaluating the box graph of Fig. 5, to constrain $\vert V_{td}
\vert$ and hence $\rho$ and $\eta$.  An analogous calculation to the one
I sketched for the Kaon system in Sec. IV.C gives \cite{IL} \cite{BSS}
\be x_d = \tau_{B_{d}} (\Delta m)_{B_{d}}{ = \tau_{B_{d}}} \frac{G^2_F
M_{B_{d}}}{6 \pi^2} [B_{B_{d}} f^2_{B_{d}} \eta_B ] m^2_t f_2 (y_t)
\vert V_{td} \vert^2
\ee
Here the parameters $[B_{B_{d}} f^2_{B_{d}} \eta_B]$ are the
counterparts in the $B_d$ system of $[B_k f^2_K \eta_2]$ in the Kaon
system .  Because the $b$ quark is heavy, one expects that the vacuum
insertion approximation should work very well, so that $B_{B_{d}} \simeq
1$.  However, now in contrast to what happens in the Kaon system, the
$B_d$ decay constant $f_{B_{d}}$ is not measured.  Nevertheless, this
parameter can be computed also by using lattice QCD methods.  Taking
into account of the short distance QCD correction factor $\eta_B \simeq
0.85$ \cite{JSH}, the best value for the factor $ \sqrt{{B_{B_{d}}}
\eta_B}~  f_{B_{d}}$ which one obtains from lattice QCD computations is
\cite{Martinelli}
\be
\sqrt{B_{B_{d}} \eta_B}~ f_{B_{d}} = (200 \pm 35)~ MeV~~.
\ee
Using the above and approximating $f_2 (y_t)$ in the same way as was
done in the Kaon system [c.f. Eq. (82)] \cite{BSS} one has that
\be
x_d = [ 0.44 \pm 0.15] A^2 [\eta^2 + (1 - \rho)^2]
(\frac{m_t}{M_W})^{1.6}~~,
\ee
where the error is essentially that coming from Eq. (96).  Using the
experimental values for $x_d$ and $A$ this gives the third constraint in
the $\rho - \eta$ plane alluded above:
\be
(1.80 \pm 0.77) = [ \eta^2 + (1 - \rho)^2] (\frac{m_t}{M_W})^{1.6}~~.
\ee

The constraints in the $\rho - \eta$ plane coming from $\vert \epsilon
\vert$, [Eq. (86)], $\vert V_{ub} \vert / \vert V_{cb} \vert$
[Eq. (87)], and from $B^0_d - {\bar{B^0_d}}$ mixing, [Eq. (98)]
are displayed in Fig.
6 for two cases: $m_t = 140~GeV$ and $m_t = 180~GeV$.
The cross hatched region in this figure uses instead of Eq. (87) the
result of Eq. (84) for $\sigma = \sqrt{\rho^2 + \eta^2}$ obtained in the
ACM model \cite{ACM}.  Obviously the overlap region allowed by our
present theoretical and experimental knowledge of $\vert \epsilon
\vert,~x_d$ and $\vert V_{ub} \vert /\vert V_{cb} \vert$ is crucially
dependent on whether one uses Eq. (87) or the more restricture ACM
result.  This is illustrated in Fig. 7.
\begin{figure}
\vspace{10cm}
\caption[]{Allowed regions in the \protect$\rho - \eta$ plane coming
from the
measurements of \protect$\vert \epsilon \vert, ~\vert V_{ub} \vert/\vert
V_{cb}\vert$ and \protect$x_d$.}
\end{figure}

\begin{figure}
\vspace{6cm}
\caption[]{Allowed region in the \protect$\rho - \eta$ plane for \protect
$m_t = 140~GeV$.  The shaded band is the result obtained by relying only
on the ACM model.}
\end{figure}

Unfortunately, even assuming $\eta$ to be in its most restricted range
$(\eta \simeq 0.2 - 0.3)$, is not sufficient to allow for a sharp
prediction for $\epsilon^{\prime}/\epsilon$.  This arises principally
from other theoretical uncertainties incurred
in estimating the hadronic matrix elements of operators
which contribute to $\epsilon^{\prime}$.  Nevertheless, considerable
progress has been made recently in trying to tackle this question,
notably by groups in Rome \cite{Rome} and Munich \cite{Munich}
who have calculated the expectations for $\epsilon^{\prime}$ at next to
leading order and then tried to estimate the relevant matrix elements.
Because these calculations are highly technical, I will limit myself
here to give a more qualitative overview of the results obtained.
\pagebreak

As I discussed earlier, the
ratio $\epsilon^{\prime}/\epsilon$ - which is essentially the same as $Re
{}~\epsilon^{\prime}/\epsilon$ - gets contribution from two kinds of
operators:  $\Delta I = 1/2$ operators and $\Delta I = 3/2$ operators.
The former contributions are induced by gluonic Penguins and thus are of
$0(\alpha_s)$.  However, since they enter in the amplitude $Im A_0$,
the $\Delta I = 1/2$ operators are affected by the whole $\Delta I
= 1/2$ suppression factor of $Re A_2/Re A_0 \simeq
1/20$ [c.f. Eq. (42)]. On the other hand,
the $\Delta I = 3/2$ contributions arise from
electroweak Penguin diagrams and thus are only of $0 (\alpha)$.
However, $Im A_2$ is measured relative to $Re A_2$ and so, effectively,
it is
not suppressed by the $\Delta I = 1/2$ factor of $Re
A_2/Re A_0$.  Furthermore, these contributions grow quadratically with $m_t$,
 while
those
of the gluonic Penguins only depends on $m_t$ as $\ell n~ m_t$.

The structure of the result of the
calculations of $\epsilon^{\prime}/\epsilon$ can be written as
follows \cite{Rome} \cite{Munich}:
\be
\frac{\epsilon^{\prime}}{\epsilon} = A^2 \eta \left\{
<2 \pi ; I = 0 \vert \sum_i C_i 0_i \vert K^0 > ( 1 - \Omega_I
)-< 2 \pi ; I = 2 \vert \sum_i {\tilde{C}}_i {\tilde{0}}_i \vert K^0
>\right\}
\ee
Here $0_i$ and ${\tilde{0}}_i$ are, respectively, $\Delta I = 1/2$ and
$\Delta I = 3/2$ operators and their coefficients $C_i$ and
${\tilde{C}}_i$ have the characteristic dependence on $\alpha_s \ell n~
m_t$ and $\alpha~ m_t^2$ alluded to above.  $\Omega_I$ is a correction to
the $\Delta I = 1/2$ contribution, which arises as a result of isospin
violation through $\pi^0 - \eta$ mixing \cite{BG} and is estimated to be
$\Omega_I = 0.25 \pm 0.10$.  Note also
in the above the characteristic
CKM dependence of $\epsilon^{\prime}$ - for a fixed given $\epsilon$ -
on the CKM parameters $A^2 \eta$.  Thus, even if the hadronic matrix
elements were perfectly known, present uncertainties in $A$ and $\eta$
would give about a $50\%$ uncertainty in $\epsilon^{\prime}/\epsilon$ - a bit
less if one could restrict $\eta$ to the ACM range.

It is difficult to extract directly from the work of the Rome
\cite{Rome} and Munich \cite{Munich} groups a value for the
coefficient of $A^2 \eta$, typifying the hadronic uncertainty in
$\epsilon^{\prime}/\epsilon$.  Nevertheless, from these papers, more to
get a feeling for the expectations than as a hard and fast result, I
infer the following.  For moderate $m_t$ - say $m_t = 140~GeV$ - gluonic
Penguins dominate.  Here the uncertainty in the matrix elements is more
under control, perhaps being only of order $30\%$.  A representative
prediction for $m_t$ in this range appears to be
\be
\frac{\epsilon^{\prime}}{\epsilon} = (11 \pm 4) \times 10^{-4} A^2
\eta~~~(m_t = 140~ GeV)~~.
\ee
For larger $m_t$ values $(m_t \simeq 200~ GeV)$ electroweak Penguins begin
to be important and they tend to cancel the contributions of the gluonic
Penguins.  The error in the matrix element estimation remains similar
in magnitude, but the central value for the overall
contribution is considerably reduced.  A representative prediction for
$m_t = 200~GeV$ is, perhaps,
\be
\frac{\epsilon^{\prime}}{\epsilon} = (3 \pm 4 ) \times 10^{-4} A^2 \eta
{}~~~(m_t = 200~GeV)~~.
\ee

If one takes the above numbers at face value, one sees that, with the
present range of $\eta$ allowed by the information on $\vert \epsilon
\vert, x_d$ and $\vert V_{ub}\vert/\vert  V_{cb} \vert$, the CKM
paradigm tends to favor rather small values for
$\epsilon^{\prime}/\epsilon$.  Typically, perhaps,
$\epsilon^{\prime}/\epsilon \simeq 4 \times 10^{-4}$, with a theory
error probably of the same order!  Such small values for
$\epsilon^{\prime}/\epsilon$ are perfectly compatible with the results
obtained by the E731 collaboration \cite{EZ31}, but are a bit difficult
to reconcile with the results of NA31 \cite{NA31}.

\subsection{Other CP Violating Processes Involving Kaons}

The forthcoming round of high precision experiments at CERN and Fermilab (as
well as at the Frascati $\Phi$-factory, which is presently under construction),
should measure $\epsilon^\prime/\epsilon$ to an accuracy of order
$\delta(\epsilon^\prime/\epsilon) \sim 10^{-4}$.  This should be sufficient to
establish that there exists indeed a $\Delta S=1$ CP
violating \linebreak phase.\footnote{It is perhaps worthwhile re-emphasizing
the
 obvious here.  Namely
that a measurement of $\epsilon$ itself is not a proof of the CKM paradigm at
all, even though qualitatively the magnitude of $\epsilon$ agrees with that of
the expectations of this paradigm.  In fact, as Wolfenstein \cite{Wl}
pointed out long ago,
$\epsilon$ could be purely the result of a new CP-violating $\Delta S=2$
superweak interaction, and have nothing to do at all with any $\Delta S=1$ CKM
phase.} Even so, there is a substantial effort underway to explore other
suitable processes sensitive to {\bf direct} CP violation (i.e. $\Delta S=1$ CP
violation) in the Kaon complex.

An obvious way to establish the existence of direct CP violation is afforded by
$K^\pm$-decays.  Any asymmetry between the partial rates of a $K^+$
into some final state $f^+$ and a $K^-$ into the state $f^-$ would be a
signal of direct CP violation, since $K^+ \leftrightarrow K^-$
mixing is forbidden by charge conservation!  However, estimates of the expected
magnitude of the asymmetry between the rates for charged kaons into a variety
of final states are quite small and prospects for detecting direct CP violation
this way are rather bleak.  Another possibility, which would be nice but again
is difficult experimentally, would be to try to measure the equivalent of the
$\epsilon^\prime$ parameter for $K_s$ decays, since there is some
expectation that this parameter is perhaps somewhat larger than the usual
$\epsilon^\prime$ parameters \cite{threezero}.
However, at the Frascati $\Phi$
factory it will not even be possible to measure $\eta_{000}$ at the expected
level.  ($\eta_{000}\simeq \eta_{+-}$), never mind getting to the level of
$\epsilon^\prime_{000}$!

An equally challenging possibility, but again one that is quite interesting
theoretically, is connected to the observations that, if $K_L$ were a pure CP
odd eigenstate, then the process

\be
K_L\rightarrow \pi^0J^*,
\ee
where $J^*$ is a (virtual) spin-one state, is forbidden by CP.  Thus,
provided they are dominated by an effective spin-one
state, the processes $K_L\rightarrow \pi^0 \ell^+\ell^-$ or $K_L\rightarrow
\pi^0\nu \bar \nu$ could be used to test CP.\footnote{In the case of $K_L$
going to charged leptons, the two-photon contribution is estimated to be small,
so effectively this process is dominated by a virtual spin-one state
\cite{DibGil}.  This is clearly not a problem for the decay into neutrino
pairs.}  Of course, as in the decay of $K_L\rightarrow \pi^+\pi^-$ one must
still separate out in these decays the direct CP effects from those coming from
mixing.  However, here the situation is different than in the $2\pi$ case.  It
turns out that for the decay $K_L\rightarrow \pi^0\ell^+\ell^-$ both the
direct and the
mixing contributions are roughly of the same order of magnitude \cite{DibGil},
while for the $K_L\rightarrow \pi^0 \nu\bar \nu$ decay
the mixing effects are in fact
totally negligible \cite{Lichten}!  Unfortunately, both of these decays are
second order weak processes and therefore the expected branching ratios are
tremendously small.  In the forthcoming round of experiments at Fermilab, one
will get near the range of interest for the $K_L \rightarrow \pi^0\ell^+\ell^-$
decay but no real test of CP violation will ensue.

In Table 2, which is adapted from the recent review of Winstein and
Wolfenstein \cite{WW}, I summarize the expectations for various other CP
violating processes in the Kaon sector, along with the accuracy one may hope to
reach in forthcoming experiments.  It should be clear from this Table that to
test for direct CP violation in this way will require yet a further round of
experiments, beyond those now planned.

\begin{table}
\caption{Expectation and Prospect for Various Kaon CP Violation Experiments}
\begin{tabular}{c|c|c}
Process & CKM Expectation & Experimental Prospect \\
\hline
& & \\
\begin{tabular}{c}
$K_s \rightarrow 3 \pi^0$ \\
$\eta_{000} = \epsilon + \epsilon_{000}^\prime$
\end{tabular}
&
$ {\displaystyle\epsilon^\prime_{000}\over \displaystyle \epsilon}
  \sim 10^{-2}$
&
\begin{tabular}{c}
$\delta\eta_{000} \sim 5 \times 10^{-3}$ \\
($\Phi$ Factory)
\end{tabular}
\\
& & \\
\hline
& & \\
\begin{tabular}{c}
$K^\pm \rightarrow \pi^\pm \pi^\pm \pi^\mp$ \\
$\Delta\Gamma = (\Gamma^+ - \Gamma^-) / (\Gamma^+ + \Gamma^-)$ \\
$\Delta g = (g^+ - g^-) / (g^+ + g^-)$
\end{tabular}
&
\begin{tabular}{c}
$\Delta\Gamma < 10^{-6}$ \\
$\Delta g < 10^{-4}$ \\
(Dalitz plot asymmetry)
\end{tabular}
&
\begin{tabular}{c}
$\delta(\Delta\Gamma) \sim 5 \times 10^{-5}$ \\
$\delta(\Delta g) \sim 5 \times 10^{-4}$ \\
($\Phi$ Factory)
\end{tabular}
\\
& & \\
\hline
& & \\
\begin{tabular}{c}
$K^\pm \rightarrow \pi^\pm \pi^0 \gamma$ \\
$\Delta\Gamma = (\Gamma^+ - \Gamma^-) / (\Gamma^+ + \Gamma^-)$ \\
\end{tabular}
&
$\Delta\Gamma < 10^{-4} - 10^{-5}$
&
\begin{tabular}{c}
$\delta(\Delta\Gamma) \sim 2 \times 10^{-3}$ \\
($\Phi$ Factory)
\end{tabular}
\\
& & \\
\hline
& & \\
\begin{tabular}{c}
$K_L \rightarrow \pi^0 \ell^+ \ell^-$ \\
$B(K_L \rightarrow \pi^0 \ell^+ \ell^-)$
\end{tabular}
&
\begin{tabular}{c}
$B_{\rm direct} \sim 10^{-11} - 10^{-13}$ \\
(depends on $m_t$ and $V_{CKM}$)
\end{tabular}
&
\begin{tabular}{c}
$B < 7 \times 10^{-11}$ \\
(E832 FNAL)
\end{tabular}
\\
& & \\
\hline
& & \\
\begin{tabular}{c}
$K_L \rightarrow \pi^0 \nu \bar\nu$ \\
$B(K_L \rightarrow \pi^0 \nu \bar\nu)$
\end{tabular}
&
\begin{tabular}{c}
$B \sim 10^{-11} - 10^{-12}$ \\
(depends on $m_t$ and $V_{CKM}$)
\end{tabular}
&
\begin{tabular}{c}
$B \sim 10^{-8}$ \\
(E832 FNAL)
\end{tabular}
\\
& & \\
\hline
\end{tabular}
\end{table}

\section{Electric Dipole Moments}

The expectation for the electric dipole moment (edm) of the neutron and the
electron are that these quantities are extremely small in the CKM model.  For
the electron case, one needs to invoke mixing in the leptonic sector and this
vanishes in the limit that the neutrinos are degenerate in mass.  So the edm
for the electron is truly vanishingly small in the standard paradigm.  It turns
out that the edm for the neutron is also very suppressed.  First of all, it is
easy to see that no edm at the quark level appears at one-loop, since these
graphs involves either $V_{di}^*V_{di}$ or $V_{ui}^*V_{ui}$ and all phase
information is lost.  It turns out that the quark edm (and hence the neutron
electric dipole moment) also vanishes at the two-loop level\cite{Shabelin}.
There is no simple explanation, as far as I know, for this result.  Indeed
individual two-loop graphs are non-vanishing, but the sum of all the graphs
contributing to the edm vanishes.  There is no full calculation of the edm at
three-loops, but one can get an order of magnitude estimate of the effect by
simply festooning the two-loop graphs with gluons.  This gives \cite{Nir}

\be
d_n\sim em_d \frac{G_F\alpha\alpha_s}{\pi}
  \frac{m_t^2m_s^2}{M_W^4} \lambda^6A^2\sigma \sin\delta \sim 10^{-33}ecm~~.
\ee

Somewhat larger estimates than the above have been obtained by considering, for
instance, the contributions to $d_n$ which arise from two-quark graphs in a
neutron \cite{Nanopoulos}.  This notwithstanding, it is clear that with CKM
paradigm an edm for the neutron above, say, $10^{-31}-10^{-32}$ ecm is very
unlikely.  Experimentally, such values are a factor of $O(10^6)$ below the
present bounds and are essentially unreachable.  This is both good and bad
news.  The good news is that by continuing to probe for a non-zero edm for the
neutron (or the electron) one is assured that any {\bf positive} result will
necessarily be a signal of new physics.  The bad news is that there is no real
assurance that any such signal will be found.

In looking for signals for new physics in connection with the neutron electric
dipole moment, it is useful to examine which effective operators can give rise
to an edm.
There are a number of effective QCD operators which can contribute to the
electric dipole moment for the neutron.  The most famous of these is the CP odd
two-gluon operator

\be
{\cal{L}}_{CP\,viol}=\bar\theta \frac{\alpha_s}{8\pi}
  G_a^{\mu\nu}\tilde G_{a_{\mu\nu}}~.
\ee
This operator arises naturally in QCD as a result of the non-trivial nature of
the QCD vacuum\cite{vacuum} and $\bar\theta$ represents a combination of the
vacuum angle contribution and that from the quark mass matrix
($\bar\theta=\theta+ {\rm Arg}\,\, det\,\, M$).  The electric dipole moment for
the neutron coming from such a term is enormous, unless $\bar\theta$ is very
small.  One can estimate\cite{Balavi} for $d_n$ a value:

\be
d_n\sim \bar\theta\biggl(\frac{m_d}{M_n^2}\biggr)
  ecm \,\, \sim 4\times 10^{-16} \bar\theta \,\, ecm~.
\ee
To agree with the experimental bound given in Eq. (10), $\bar\theta$ has to be
vanishingly small $(\bar\theta \leq 10^{-10})$!  Why this should be so, is the
strong CP problem.  The only sensible solution, to my mind, of this
conundrum
is that actually $\bar\theta \equiv 0$ for dynamical reasons \cite{1a}.  In
this case, of course, the operator in Eq. (104) does not contribute to $d_n$.

Even if $\bar\theta=0$ dynamically, one can always get a contribution to the
electric dipole moment of the neutron from an induced
CP odd three-gluon operator

\be
{\cal{L}}_{CP\, viol} = \frac{1}{\Lambda^2} f_{abc} G_a^{\mu\nu}
  G_{b\nu}^\alpha \tilde G_{c\alpha\mu} ~.
\ee
The scale $\Lambda$ for the CKM model is effectively extremely large because
the GIM mechanism \cite{GIM} introduces high powers of $(m_q/M_W)^n$.  This is
not necessarily so in other models and one can get a sizable edm from the
operator in Eq. (106).  Typically

\be
d_n \sim \frac{eM_n}{\pi\Lambda^2}
\ee
and edm's in the neighborhood of $10^{-26}$ ecm are perfectly plausible in a
variety of models\cite{Weinberg}.  For these reasons, it seems very sensible to
continue to probe experimentally as hard as one can for a non-vanishing edm.

\section{CP Violation in the B System}

In decays of $B$ mesons CP asymmetries do not have to suffer from the
family mixing suppression factors one encounters in the Kaon sector.
After all, $B_d$ (or $B_s$) mesons contain already quarks from the third
and first (or second) generation and their decay by-products can easily
involve states containing quarks from yet another generation.  The
presence of all three generations in the relevant decay amplitudes
serves to remove, for certain processes, the family mixing suppression
factors which arose through virtual intermediate states in the Kaon
system.  As we shall see, the best place to see directly the CP
violating phase of the CKM paradigm is by studying CP violating
asymmetries in $B$ decays to {\bf CP-self conjugate states} $f$.  These
are states which have the property that $\bar{f} = \pm f$.  Before
showing why this is so, however, it is necessary to develop a bit of
formalism and detail a certain amount of information related to $B$
decays.

Because $B$ mesons are quite heavy, they decay into a large number of
distinct channels.  Thus, as we discussed in Sec. IV.D, in contrast
to what happens in the Kaon system, one expects that in the neutral $B$
sector $\Gamma_+ \simeq \Gamma_-$. Hence
\be
\Gamma = \frac{1}{2} (\Gamma_+ + \Gamma_-) \gg \frac{1}{2} (\Gamma_+ -
\Gamma_-)~~.
\ee
Furthermore, for both the $B^0_d$ and $B^0_s$ systems the mass difference
between the eigenstates $\Delta m$ is much less than the average mass:
\be
\frac{1}{2} (m_+  + m_-) \gg \frac{1}{2} \Delta m = \frac{1}{2} (m_+ -
m_-)~~.
\ee
However, as I remarked upon earlier, experimentally $\Delta m$ is quite
comparable to $\Gamma$ for the case of the $B_d$ states \cite{Cassel}
[cf Eq. (94)]
\be
\biggl(\frac{\Delta m}{\Gamma}\biggr)_{B_{d}} \equiv x_d = 0.64 \pm 0.08
\ee
Although there is no measurement to date of $\Delta m/\Gamma$ for the
$B_s$ system, one expects that also this quantity be of $0(1)$.
As a result, the time evolution of a $\vert (B_s){\rm phys} >$ state
will be also given by an equation similar to Eq. (90) with $B_d
\leftrightarrow  B_s$.  In what follows, therefore, I shall not
distinguish explicitly between the $B_d$ and $B_s$ cases - until that is
needed - and write simply
\bea
\vert  B_{\rm phys} (t) > = & e^{ -i m_B t} e^{- \frac{1}{2} \Gamma_B t}
\Biggl\{ \cos \frac{\Delta m_B t}{2} \vert B > + \nonumber \\
& + i \eta_B e^{i \Phi_B} \sin \frac{\Delta m_B t}{2} \vert \bar{B} >
\Biggr\}~~.
\eea
\bea
\vert {\bar{B}}_{\rm phys} (t) > = e^{- i m_B t} e^{- \frac{1}{2}\Gamma_B t}
&\Biggl\{ \frac{i {\bar{e}}^{- i \Phi_B}}{\eta_B} \sin \frac{\Delta m_B
t}{2} \vert B > \nonumber \\
& + \cos \frac{\Delta m_Bt}{2} \vert \bar{B} > \Biggr\}~~.
\eea

For the $B$ system, the off diagonal matrix element $M_{12} \sim m_t^2$,
while the off-diagonal width $\Gamma_{12} \sim m_c^2$ (since the decays
$b \to c$ dominate) \cite{Punk}.  Thus using Eq. (33) one expects that
\be
\eta_B e^{i \Phi_B} = \frac{1 - \epsilon_B}{1 + \epsilon_B } =
\Biggl[\frac{(M_{12}^{\ast} - \frac{1}{2} \Gamma_{12}^{\ast})^{1/2}}{(M_{12} -
\frac{i}{2} \Gamma_{12})^{1/2}} \Biggr]_B \simeq
\Biggl[\frac{M_{12}^{\ast}}{M_{12}} \Biggr]^{1/2} \equiv e^{- i \Phi_M}~~.
\ee
That is, $\eta_B \simeq 1$, with the phase $\Phi_B$ being essentially
the negative of the phase $\Phi_M$ of the appropriate mixing matrix
$M_{12}$.  For the $B$ system  this mixing matrix is dominantly given by
computing a box graph with a $t$ quark in the loop [cf Fig. 5 for the
$B_d$ case].  It is easy to see that for $q = \{ d, s\}$
\be
(M_{12})_{B_{q}} \sim [V_{tb} V^{\ast}_{tq}]^2 \sim [V^{\ast}_{tq}]^2~~,
\ee
since $V_{tb} \simeq 1$ to leading order in $\lambda$.  Using Eq. (26)
for the CKM matrix one observes that, to leading order in $\lambda,
V_{ts}$ is real while $V_{td}$ has a complex phase.  Writing
\be
V_{td} = \vert V_{td} \vert e^{- i \beta}~~,
\ee
one secures the result
\bea
(M_{12})_{B_{q}} = \vert M_{12} \vert_{B_{q}}\cdot
\left\{\begin{array}{cc}
e^{+2 i \beta}& \mbox{($q = d $)}\\
1&\mbox{($q= s$)}\end{array}\right.
\eea
Hence, in the CKM paradigm, for the neutral $B$ states one can take in
Eq. (113)
\bea
\eta_B \simeq 1~~,~~ \Phi_B \simeq \left\{ \begin{array}{cc}
- 2 \beta &\mbox{($q = d$)} \\
0 &\mbox{($ q = s$)} \end{array} \right.
\eea

\subsection{CP Violation in Decays to CP-Self Conjugate States}

It is interesting to study the decays of $B_{\rm phys}$ states into
CP-self conjugate states $f$ - with $\bar{f} = \pm f$.  The ratio of the
decay amplitudes of $B$ and $\bar{B}$ to one of these states $f$ would
have unit magnitude if CP were conserved, but in general one expects that
\be
\frac{A (\bar{B} \to f)}{A (B \to f)} = \eta_f e^{i \Phi_D}~~.
\ee
Using this equation, it is straightforward to compute the (time
dependent) rate of $B_{\rm phys} (t)$ and ${\bar{B}}_{\rm phys} (t)$ to
decay into a CP self conjugate state $f$:
\bea
\Gamma (B_{\rm phys} (t) \to f) = \Gamma (B \to f)& e^{- \Gamma_B t}
\Biggl\{ \cos^2 \frac{\Delta m_Bt}{2} + \eta^2_f \sin^2 \frac{\Delta
m_B t}{2}\nonumber \\
& - \eta_f \sin (\Phi_B + \Phi_D) \sin \Delta m_B t \Biggr\}
\eea
\bea \Gamma ({\bar{B}}_{\rm phys} (t) \to f) = \Gamma (B \to f)
&e^{-\Gamma_B t} \Biggl\{ \eta^2_f \cos^2 \frac{\Delta m_B t}{2} +
\sin^2 \frac{\Delta m_B t}{2} \nonumber \\
& + \eta_f \sin (\Phi_B + \Phi_D) \sin \Delta m_B t \Biggr\}~~.
\eea

{}From CPT conservation
one can obtain a relation between the amplitudes of $A (B \to
f)$ and $A (\bar{B} \to \bar{f})$.  Since for CP self conjugate states
$A (\bar{B} \to \bar{f}) = \pm A(\bar{B} \to f)$, this relation has a
bearing on the desired ratio of Eq. (118).  In general the amplitude
$A(B \to f)$ will contain different weak amplitudes $a_i$ each
multiplied by an appropriate strong rescattering phase factor $e^{i
\delta_i}$:
\be
A(B \to f) = \sum_i a_i e^{i \delta_i}~~.
\ee
CPT conservation
gives for the corresponding amplitude for $A (\bar{B} \to \bar{f})$
\be
A(\bar{B} \to \bar{f}) = \sum_i a^{\ast}_i e^{i \delta_i}~~.
\ee
That is, one conjugates the weak amplitudes but keeps the rescattering
phases the same [cf. Eqs. (38) and (39) for the $K^0 - {\bar{K}}^0$
complex].  Whence it follows that
\be
\eta_f e^{i \Phi_D} = \pm \frac{\sum_i a^{\ast}_i e^{i \delta_i}}{\sum_i
a_i e^{i \delta_i}}~~.
\ee

There are many circumstances where one can argue dynamically that {\bf
only one} weak amplitude dominates the ratio of the $\bar{B} \to f$ to
$B \to f$ amplitudes.  In this case clearly
\be
\eta_f = \pm 1~~~~(\bar{f} = \pm f)~~.
\ee
and the formulas for the decays of $B_{\rm phys}$ and ${\bar{B}}_{\rm
phys}$ simplify considerably.  In particular, in
this case, the asymmetry between these rates is simply a measure of the
CP violating phase $\Phi_B + \Phi_D$:
\be A(t) = \frac{\Gamma (B_{\rm phys} (t) \to f) - \Gamma
({\bar{B}}_{\rm phys} (t) \to f)}{\Gamma (B_{\rm phys} (t) \to f) +
\Gamma ({\bar{B}}_{\rm phys} (t) \to f)} = (\mp) \sin (\Phi_B + \Phi_D)
\sin \Delta m_B t~~.
\ee
In contrast to the Kaon case, however, this asymmetry is not small since
the CP violating phases $\Phi_B$ and $\Phi_D$ are not suppressed by
small mixing angles,  Indeed, we just saw above that for the $B_d$ case
$\Phi_{B_{d}} = - 2 \beta$, with $- \beta$ being the phase of $V_{td}$.

The above nice formula holds for decays $B \to f$ and ${\bar{B}} \to f$
which are dominated by one weak amplitude.  In this case the
strong rescattering phases cancel, since they are common for the single
$B \to f$ and $\bar{B} \to f$ transition amplitudes.  Thus $\Phi_D$ is
purely a weak CP violating phase.  Because the $B$'s are heavy, we can
to a good approximation compute their  decays using the spectator
picture \cite{spectator}, in which the weak amplitude for a $B$ (or a
$\bar{B}$) to decay is just proportional to the corresponding weak
amplitude for its constituent $\bar{b}$ (or $b$) quark to decay.  That is
\be
\frac{A (\bar{B} \to f)}{A (B \to f )} \simeq (\pm) \frac{A (b \to q
q^{\prime} {\bar{q}}^{\prime\prime})}{A (\bar{b} \to \bar{q}
{\bar{q}}^{\prime} q^{\prime \prime})}
\ee
If one ignores Penguin effects, in the above $q = \{ u, c\}$ while
$q^{\prime}$ and $q^{\prime\prime}$ are quarks of the first two
generations with $q^{\prime} = \{ d, s\} $ and
$q^{\prime\prime} =  \{ u, c\}$. If it were not for
the presence of the CKM phase, the $b$ quark
decays amplitudes would be real.  Since according to Eq. (26), to leading
order in $\lambda,  V_{q^{\prime\prime} q^{\prime}}$ is real, one sees
that in this approximation:
\bea
\frac{A(\bar{B} \to f)}{A(B \to f)} \simeq (\pm)
\frac{V_{qb}}{V_{qb}^{\ast}} = (\pm)
\left\{\begin{array}{cc}
e^{- 2 i \delta} & \mbox{($b \to u$ {\rm transition})}\\
 1 & \mbox{($b \to c$ {\rm transition})} \end{array}\right.~~.
\eea
Thus the decay phase $\Phi_D$ either is directly related to
the CKM phase $\delta$ or vanishes!  That is,
\bea
\Phi_D \simeq \left\{\begin{array}{cc}
- 2 \delta & \mbox{($b \to u$ {\rm transition})}\\
0 & \mbox{($b \to c$ {\rm transition})}\end{array} \right.
\eea

\subsection{The Unitarity Triangle and Classes of Predictions}

When one weak amplitude dominates in the decay of neutral $B$ mesons to
CP self-conjugate states, the decays of $B_{\rm phys}(t)$ and
${\bar{B}}_{\rm phys}(t)$ into these states only differ by the sign of
the modulating factor $\sin (\Phi_B + \Phi_D) \sin \Delta m_B t$.  As a
result, the asymmetry between these rates, normalized to the sum of the
rates, measures precisely this factor [c.f. Eq. (125)].  The coefficient
of $\sin \Delta m_Bt$ is a measure of CP violation in the $B - \bar{B}$
complex.  Here, however, in contrast to what obtains in the Kaon case,
one expects this coefficient to be sizable.

Let us write:
\be
\alpha_f = \mp \sin (\Phi_{B} + \Phi_D)~~.
\ee
As we saw in the last section, in the CKM model these CP violating
phases are directly related to the phases of the complex CKM matrix
elements. Furthermore, to leading order in $\lambda$, we identified only
four possibilities for these phases, which are described by Eqs. (117)
and (128).\footnote{$\Phi_B$ and $\Phi_D$ are convention dependent and
the results given are those which follow if one uses the quark phase
convention.  However, the sum $\Phi_B + \Phi_D$ is convention
independent, as it must be.}  Thus one can divide the expectations for
the coefficients $\alpha_f$ into four different classes, depending on
whether one is dealing with $B_d$ or $B_s$ decays and depending on
whether the main weak transition involves a $b \to c$ or a $b \to u$
process.  The relevant results are displayed in Table 3, in which also
typical decay processes are identified.
\begin{table}
\caption{CKM Model expectations for the hadronic asymmetry coefficients
\protect$\alpha_f$ to CP-self conjugage states \protect$f = \pm
{\bar{f}}$}
\vspace*{.2cm}
\begin{center}
\begin{tabular}{| l | l | l |} \hline
{}~~~~~~~Process &~~~~$\alpha_f$ &{\rm Typical example} \\ \hline
$B_d$ {\rm decay}~~;~~ $b \to c $ {\rm transition}   & $\pm \sin 2
\beta$ & $B_d \to \psi K_s$ \\ \hline
$B_d$ {\rm decay}~~;~~ $b \to u$ {\rm transition}  & $\pm \sin 2 (\beta +
\delta)$ & $\protect B_d \to \pi^+ \pi^-$\\ \hline
$B_s$ {\rm decay}~~;~~$ b \to c~ {\rm transition}$  &~~~~ $0$ & $B_s \to \psi
\phi$
\\ \hline
$B_s$ {\rm decay}~~;~~$b \to u~{\rm transition}$ & $\pm \sin 2 \delta$ & $
B_s \to \pi^0 K_s$ \\ \hline
\end{tabular}
\end{center}
\end{table}

There is a very nice interpretation of the above results, whose origins
can be traced to Bjorken\cite{Bj}, although the realization that the CKM
model has the simple set of results of Table 2 probably predates this
interpretation \cite{unitary}.  The angles which enter the different
asymmetries in Table 3 $\{ \delta, \beta, \delta + \beta\}$ are angles
of a triangle intimately connected with the CKM matrix.  Indeed, the
existence of this (approximate) triangle is just a simple reflection of
the unitary of $V_{CKM}$.  Consider for this purpose the unitarity
relation for the $bd$ matrix element of $V^{\dagger}_{CKM} V_{CKM}$.  Since
 $V_{CKM}$
is unitary one has
\be
V^{\ast}_{ub}  V_{ud} + V^{\ast}_{cb} V_{cd} + V^{\ast}_{tb} V_{td}
= 0~~.
\ee
The above, using Eq. (26), reduces to leading order in $\lambda$ to
\be
V^{\ast}_{ub} + V_{td} \simeq \lambda V^{\ast}_{cb}
\ee
{}~~~~or
\be
\vert V_{ub} \vert e^{i \delta} + \vert V_{td} \vert e^{- i \beta}
\simeq A \lambda^3~~.
\ee
The above describes a triangle in the complex plane,
two of whose angles are $\delta$ and
$\beta$, as shown in Fig. 8a.  Using that
\bea
V_{ub} =& A \lambda^3 \sigma e^{- i \delta}~ =\hspace*{1.0em} A \lambda^3 (\rho
 - i
\eta)\nonumber \\
V_{td} =& A \lambda^3 (1 - \sigma e^{i \delta} )~ =\hspace*{1.0em} A \lambda^3
(
 1 -
\rho - i \eta) ~~,
\eea
and scaling Eq. (132) by $A \lambda^3$, one sees that the ``unitarity"
triangle of Fig. 8a becomes the triangle in the $\rho - \eta$ plane
shown in Fig. 8b.
\begin{figure}
\vspace*{8cm}
\caption[]{a) Unitarity triangle showing the angles \protect$\beta$ and
\protect$\delta$; b) This same triangle rescaled, shown in the
\protect$\rho - \eta$ plane.}
\end{figure}

As can be seen from Fig. 8b, the tip of the unitarity triangle is the
point $\{\rho , \eta \}$.  As we discussed in Sec. IV.D, this point is
{\bf not} uniquely determined by the present measurements of Kaon CP
violation, $V_{ub}/V_{cb}$ and $ B - \bar{B}$ mixing, mostly due to
uncertainties in the theory.  From our analysis of these measurements,
the allowed region in the $\eta - \rho$ plane is that shown in Fig. 7.
Even restricting oneselves to the small strip favored by the
ACM model, one still has a variety of possible unitarity triangles
and two of these
are shown in Fig. 9.  Even given this uncertainty, in all
cases, it appears that the important angles $\delta$ and $\beta$, and
their complement $(\delta + \beta)$, which determine $\alpha_f$ are
sizable. Thus in the CKM model one expects CP violating
asymmetries $\alpha_f$ which are at the level of $10\%$ or so, rather
than at the level of $10^{-3}$ which is what is
observed in the neutral Kaon system.
\begin{figure}
\vspace*{10cm}
\caption[]{Examples of allowed unitarity triangles which follow from our
analysis of the Cabibbo Kobayashi Maskawa matrix.}
\end{figure}

There are a number of analyses in the literature for the expectations of
the CKM model for the asymmetry coefficients
$\alpha_f$~\cite{expectations}.  I give in Fig. 10, as an example, the
results of Nir \cite{Nir} in which the allowed region in the $\sin 2
\beta-\sin 2 (\beta + \delta)$ plane is shown.  As can be seen from
this figure, there is always a sizable asymmetry coefficient $(\sin 2
\beta)$ for $B_d$ decays to CP self-conjugate states involving a $b \to
c$ transition.  Thus the decays $B_d \to \psi K_s~~, {\bar{B}}_d \to
\psi K_s$, prototypical of these processes, should be prime
candidates for observing CP violation in the neutral $B$ system.

The above nice results are predicated on having only one weak amplitude
dominate the decays in question.  It is clearly important to examine the
limits of validity of this approximation.  This is particularly
necessary for decays into CP self-conjugate states, since in these cases the
quark decays $b \to q q^{\prime} {\bar{q}}^{\prime\prime}$ always have
$q^{\prime\prime} = q$.  Thus for these decays, associated with the quark
decay amplitude there {\bf always} exists also a Penguin amplitude $b
\to q^{\prime} ( q {\bar{q}})$.  This is shown, schematically, in Fig.
11.
\begin{figure}
\vspace*{8cm}
\caption[]{Allowed values in the \protect$\sin 2 \beta - \sin 2 (\beta +
\delta)$ plane, according to the analysis of Nir \protect{\cite{Nir}}.}
\end{figure}

\begin{figure}
\vspace*{8cm}
\caption[]{Decay and Penguin amplitudes entering in the processes
\protect$b \to q^{\prime}(q \bar{q})$.}
\end{figure}
Let us examine what corrections the Penguin contributions of Fig. 11 give to
the asymmetry coefficients $\alpha_f$.  Because one now does not have a single
weak amplitude, the simple relation (124) no longer obtains and one must use
the general formula (123) for the ratio of the decay amplitudes $A(\bar
B\rightarrow f)$ to $A(B\rightarrow f)$.  In particular, if we denote the quark
decay amplitude by $A_q$ and the Penguin amplitude by $A_P$, one has

\be
\frac{A(\bar B\rightarrow f)}{A(B\rightarrow f)} = \pm
\frac{A_q^*e^{i\delta_q}+A^*_Pe^{i\delta_P}}
  {A_q e^{i\delta_q}+A_P e^{i\delta_P}}~~,
\ee
where $\delta_q$ and $\delta_P$ are the rescattering phases corresponding to
the quark
decay and the Penguin amplitudes, respectively.  In general, these strong
phases now no longer cancel and the measured asymmetry is no longer simply
related to the unitarity triangle phases.

Fortunately, a careful analysis by various people \cite{Penguin analysis} has
shown that, in most instances, the Penguin contributions are quite negligible.
For instance, for the important process $B_d\rightarrow \psi K_S$ the relevant
Penguin graph involves a $b\rightarrow s$ Penguin.  This graph, like the
associated quark decay amplitude which involves a $b\rightarrow c$ transition,
is real to leading order in $\lambda$.  Hence, in this case both $A_q$ and
$A_P$ are real and the ratio of the amplitudes in Eq. (134)--even taking into
account the Penguin effects--is again approximately $\pm 1$.  Typical estimates
of the uncertainty in $\alpha_f$ due to the presence of Penguin graphs are
given
in \cite{Penguin analysis} and are of the order of
$\delta\alpha_f/\alpha_f\leq 1\%$ for the process $B_d\rightarrow \psi K_S$ and
are of order of $\delta\alpha_f/\alpha_f \leq 10\%$ for the processes
$B_d\rightarrow \pi^+\pi^-$.

\subsection{Other CP Violating Asymmetries in the Neutral B Systems}

The holy grail for testing the CKM model is the measurement of the CP violating
asymmetries of neutral $B$ decays to CP self-conjugate states.  As we just
discussed, these asymmetries afford a direct way to check the unitarity of the
mixing matrix through the determination of the angles in the unitarity
triangles.  However, even though one expects in general rather large
asymmetries, these measurements are far from trivial experimentally.  What
makes them difficult is that the branching ratios for the various exclusive
decays $B\rightarrow f$ one wants to study are small and, furthermore, to
obtain
a value for $\alpha_f$ one needs to know how the initial state was ``born''.
The necessity to tag the events for these purposes, along with the small
exclusive branching ratios, require samples of the order of $10^7-10^8$ B's
for a good measurement.  Such samples will become available in the future in
the $e^+e^-$ B factories now projected in the USA and Japan.  Samples even
larger than these are already being produced yearly at the Tevatron Collider.
Here, however, the challenge is to extract the signal from the background.

Given these practical difficulties, it is worthwhile contemplating whether
there are other clean tests of the CKM paradigm, besides those afforded by
studying B decays to CP self-conjugate states.  It turns out that, provided one
can do a certain amount of {\bf spin tagging}, there are a number of other
measurements in the neutral B system which are as theoretically pristine as
those
we discussed in the previous section.  Some of the ideas behind these alternate
tests were first discussed in a paper by Kayser {\it et al.} \cite{KKPS} and I
will use two of the examples considered in this paper to illustrate the
general principles.  More detailed discussions can be found in \cite{SpinTag}.

A nice illustration of the complications which spin introduces when
discussing CP violating asymmetries is provided by the decay $B_d\rightarrow
p\bar p$.  Even though this final state is a CP self-conjugate state, it is not
quite correct that the CP-violating asymmetry of $(B_d)_{phys}$ to $p\bar p$
provides a clean test of the CKM model.  This is because the $p\bar p$ state,
in contrast to the examples considered earlier, has two possible helicities and
not just one.  Since CP flips helicity, not having a pure helicity final state
necessitates knowing something of the hadronic dynamics to extract the
desired asymmetry coefficient.  The CP eigenstates for the $p\bar p$ system
are linear combinations of the two states which are produced in $B_d$ decay,
which either have both $p$ and $\bar p$ carrying helicity $+1/2$, or both
carrying helicity $-1/2$.  That is, the CP eigenstates are

\be
|p\bar p;\,\,\pm> = \frac{1}{\sqrt{2}}\biggl\{|p+ \frac{1}{2}>
|\bar p+ \frac{1}{2}>
  \pm |p-\frac{1}{2}>|\bar p-\frac{1}{2}>\biggr\}~~.
\ee
The decay rates of $(B_d)_{phys}$ to the above states will have a
modulating factor whose constant of proportionality $\alpha_{p\bar p}$ is
entirely fixed by the CKM matrix:\footnote{Indeed, $\alpha_{p\bar
p}=-\alpha_{\pi^+\pi^-}$.}

\be
\Gamma\bigl((B_d)_{\rm phys}(t)\rightarrow p\bar p;\pm\bigr)=|A_\pm|^2
  e^{-\Gamma_Bt}\{1\pm \alpha_{p\bar p} \sin \Delta m_B t\}.
\ee
Because, in general, $|A_+|^2 \not= |A_-|^2$ the rate of $(B_d)_{\rm phys}$ to
$p\bar p$, however, depends on the dynamics.  One has

\begin{eqnarray}
\Gamma\bigl((B_d)_{\rm phys}(t)\rightarrow p\bar p\bigr)&=&
\begin{array}[t]{l}
\Gamma
  \bigl((B_d)_{\rm phys}
(t)\rightarrow p\bar p;+\bigr)
\\
+\:\: \Gamma\bigl((B_d)_{\rm phys}(t)\rightarrow p\bar p;
-\bigr)
\end{array} \nonumber \\
 &=&
\begin{array}[t]{l}
(|A_+|^2+|A_-|^2) e^{-\Gamma_B t} \\
 \times  \biggl\{1+\biggl(
\frac{\displaystyle |A_-|^2-|A_+|^2}{\displaystyle |A_-|^2+|A_+|^2}\biggr)
  \alpha_{p\bar p}\sin\Delta m_B \, t\biggr\}
\end{array}
\end{eqnarray}
Thus, unless one can determine the rates $|A_+|^2$ and $|A_-|^2$
independently, one cannot extract $\alpha_{p\bar p}$ from a measurement of the
asymmetry between the decays of $B_d$ and $\bar B_d$ to $p\bar p$.

The experimental selection of a pure CP eigenstate for the $p\bar p$ system is
not practical \cite{KKPS}.  However, this selection can be done quite readily
in other instances.  A nice example is provided by the decay of $B_d$ into
$\psi K^*$.  As with the $p\bar p$ case, here the final
state is not a pure helicity
eigenstate, since the $\psi$ and the $K^*$ can both have
either $\lambda=+1, \,\,
\lambda=0$ or $\lambda=-1$.  However, one can isolate the final state where the
$\psi$ and the $K^*$ have helicity $\lambda=0$, by studying the angular
distributions of the subsequent decay of the produced $K^*$ into $K_S\pi^0$.
If $\theta$ is the angle between the $K_S$ direction and that of the $K^*$,
in the $K^*$ rest frame, then one can show that \cite{KKPS}

\be
\frac{d\Gamma}{d \cos\theta}\bigl[(B_d)_{\rm phys}(t)\rightarrow K^*
  \psi \rightarrow K_S(\theta)\pi^0\psi\bigr]=\Gamma_0(t)+\Gamma_1(t)
  \sin^2\theta.
\ee
The rate $\Gamma_1(t)$ has a modulating factor whose coefficient is not
purely determined by the weak interactions.  However, $\Gamma_0(t)$ is
precisely $\Gamma\bigl((B_d)_{\rm phys}(t)\rightarrow K^*\psi; \,\,
\lambda=0\bigr)$ and as such it provides an alternative measurement of the
asymmetry coefficient $\alpha_{K_S\psi}$.  One has

\be
\Gamma_0(t)=\Gamma\bigl((B_d)_{\rm phys}(t)\rightarrow K^*\psi; \,\,
\lambda=0\bigr) = {\rm const}. \, e^{-\Gamma_B t}\bigl\{1-\alpha_{K_S\psi}
\sin\Delta m_B \,\, t\bigr\}.
\ee
That is, this rate has the same coefficient for the modulating factor as
that entering in the decay of $B_d\rightarrow \psi K_S$, except that the sign
is opposite since the $K^*$ has spin 1 while the $K_S$ has spin 0.

\subsection{CP Violating Asymmetries in Charged B Decays}

The decays of charged B mesons, like those of charged Kaons, can be
used to look for CP violation effects.  To observe a CP violation in $B^\pm$
decays one needs to have an interference between two amplitudes\footnote{For
the decay of neutral B's to CP self-conjugate states one also has, in effect,
two amplitudes--the actual decay amplitude of the B mesons themselves and the
amplitude for $B-\bar B$ mixing.}.  Furthermore these amplitudes must have both
a {\bf weak} and a {\bf strong} relative phase between each other to lead to an
observable effect.  In view of this, let us write for the rate of a $B^+$ to
decay to some final state $f^+$ the expression

\be
\Gamma(B^+\rightarrow f^+)=|A_1+A_2 e^{i\delta_W} e^{i\delta_S}|^2,
\ee
where $\delta_W, \,\, \delta_S$, are, respectively, the weak and strong
phase differences between the amplitudes $A_1$ and $A_2$, which otherwise are
taken to be real.  It follows then that the rate of $B^-$ decay to $f^-$ is
given by

\be
\Gamma(B^-\rightarrow f^-)=|A_1+A_2 e^{-i\delta_W} e^{i\delta_S}|^2 .
\ee
That is, the weak phase enters with the opposite sign as a result of charge
conjugation, but one retains the same final state strong rescattering phase.
{}From these formulas it follows, as advertised, that the asymmetry between
these
rates--which is a measure of CP violation--vanishes unless both $\delta_W \not=
0$ and $\delta_S \not= 0$ \footnote{One needs to have, of course, also
both $A_1$ and $A_2$ be nonvanishing.}:

\be
A_{+-}=\frac{\Gamma(B^+\rightarrow f^+)-\Gamma(B^-\rightarrow f^-)}
  {\Gamma(B^+\rightarrow f^+)+\Gamma(B^-\rightarrow f^-)}=
\frac{2A_1A_2 \sin\delta_W \sin\delta_S}{A_1^2+A_2^2+2A_1A_2
  \cos\delta_W \cos\delta_S}.
\ee

Because $A_{+-}$ specifically depends on the strong relative phase $\delta_S$,
estimates of these asymmetries always involve strong dynamics and thus are more
uncertain.  Furthermore, as one can see from the above formula, to get a
sizable asymmetry one must interfere amplitudes which are roughly of the same
order of magnitude, since if $A_1 \ll A_2$ then $A_{+-}\sim O(A_1/A_2)$.
Interestingly enough, there are certain types of $B^\pm$ decays where this
circumstance obtains.  Unfortunately, this typically happens for decays which
have rather small branching ratios and so the observation of a possible rate
asymmetry $A_{+-}$ will be difficult due to lack of statistics.

A good illustration of the above considerations is afforded by the doubly CKM
suppressed decays involving, at the quark level, the transition $b\rightarrow
u\bar us$.  Any of these decays can proceed either via a direct quark decay,
via $W$ exchange, or via a $b-s$ Penguin graph as illustrated in Fig. 12.
The direct decay graph is doubly suppressed because it involves a
$b\rightarrow u$ vertex as well as a $u\rightarrow s$ vertex.  Hence

\begin{figure}
\vspace{6cm}
\caption{Graphs contributing to the process \protect$b\rightarrow u\bar us$:
          (a) direct decay graph; (b)\protect$b-s$ Penguin graph.}
\end{figure}

\be
A_{direct \,\, decay} \sim O(\lambda^4) .
\ee
The $b-s$ Penguin graph, since it is dominated by $t$ and $c$ intermediate
states is only of $O(\lambda^2)$, but of course it is suppressed below this
level because of the presence of the gluon.  This suppression can make it
effectively to be of $O(\lambda^4)$ also

\be
A_{Penguin} \sim [\rm Penguin \,\, suppr.] O(\lambda^2) \sim O(\lambda^4).
\ee

Since only $V_{ub}$ has a nontrivial phase and the $u$ quark contribution in
the
Penguin graph is totally negligible,, one sees that the two amplitudes in Fig.
12 have a relative weak phase between them.  Indeed, in this case
$\delta_W \simeq \delta$, the CKM phase itself.  For any given final state, say
$B^+\rightarrow K^+\rho^0$, one also expects that the Penguin amplitude and the
direct decay amplitude give rise to different rescattering phases.
In fact, one can twist the Penguin graph involving the $c$ quark intermediate
state so that it can mimic a $D_s\bar D$ state.\cite{Pchia}  So naively one
might expect that the Penguin amplitude should carry the strong phase
associated with the rescattering process $D_s\bar D\rightarrow K^+\rho^0$,
while
the decay amplitude would not have such a rescattering phase.

Because of the difficulties of computing the strong rescattering phases,
the estimates for the
asymmetry $A_{+-}$ expected for the decays $B^\pm\rightarrow K^\pm\rho^o$
found in the literature vary
considerably, from less than a percent to over 10\% \cite{Bcharged}.  Because
this process is suppressed by higher powers of CKM mixing angles, one expects
branching ratios for these decays below $10^{-5}$, making the detection of
even a substantial asymmetry very difficult.  Thus the prospects of finding CP
violation in charged B decays are not very promising.  Asymmetries in the rates
of charged B decay should certainly be
looked for, because they are in many ways simpler experimentally (no need to
tag, mostly charged tracks in the final state, etc.).  However, their
interpretation necessarily
will need strong interaction input and one will have to be very lucky to see a
signal at all!

\section{Acknowledgements}

I am extremely grateful to Jogesh Pati for having invited me to lecture in
Puri.  This proved to be a most enjoyable experience, particularly because of
the enthusiastic response which the students at the school gave to my lectures.
I am also extraordinarily grateful to J. Maharana who managed to vector me in
and out of Puri in the midst of the Indian Airlines' strike and ongoing civil
disturbances with no glitch at all--a remarkable (and probably unrepeatable)
performance.  Bravo!  Last but not least, I would like to thank Liviana Forza
for her many kindnesses in Puri and her patience while waiting for my
manuscript to materialize.

This work was supported in part by the US Department of Energy under Contract
No. DE-FG03-91ER40662.

\end{document}